\newcommand{\noisev}{ {n} }

\newcommand{\signalv}{ {s} }

\newcommand{\ijp}{ i-j\,\textrm{pairs}}
\newcommand{\Ra} {{\,^{i}\negthinspace R}}
\newcommand{\Rb} {{\,^{j}\negthinspace R}}
\newcommand{\sumtim} {{\sum_{k=1}^{m}}}

\newcommand{\A} { {\,^{i}\negthinspace}}
\newcommand{\B} { {\,^{j}\negthinspace}}
\newcommand{\AB} { {\,^{ij}\negthinspace}}
\newcommand{\Lag} {{\cal L}}
\newcommand{\ntel} { N_{\rm tel}}
\newcommand{\gain} { {\cal G}}

\newcommand{\EQ}[1] {equation~(\ref{#1})}
\newcommand{\SEC}[1] {Section~\ref{#1}}
\newcommand{\APP}[1] {Appendix~\ref{#1}}
\newcommand{\FIG}[1] {Figure~\ref{#1}}
\newcommand{\TAB}[1] {Table~\ref{#1}}
\newcommand{\D} {\mathrm{d}}

\documentclass[useAMS, usenatbib, twocolumn]{mn2e}
\usepackage{graphicx}

\usepackage{empheq, amsmath}
\title[The optimal schedule for pulsar timing array observations]{The optimal
schedule for pulsar timing array observations}
\author[K.~J.~Lee et al.]{
K.~J.~Lee $^{1,2}$\thanks{Email: kjlee@mpifr-bonn.mpg.de},
C.~G.~Bassa$^{2}$, G.~H.~Janssen$^{2}$,
R.~Karuppusamy$^{1,2}$, \newauthor \
M.~Kramer$^{1,2}$,
R.~Smits$^{2,3}$
and
B.~W.~Stappers$^{2}$
\\
$^1${Max-Planck-Institut f\"ur Radioastronomie, Auf dem H\"ugel 69,
D-53121 Bonn, Germany} \\
$^2${Jodrell Bank Centre for Astrophysics, University of Manchester,
Manchester M13 9PL, UK} \\
$^3${Stichting ASTRON, Postbus 2, 7990 AA Dwingeloo, The Netherlands} \\
}

\begin{document} \date{\today}
\pagerange{\pageref{firstpage}--\pageref{lastpage}} \pubyear{2011}
\maketitle \label{firstpage}

\begin{abstract} 
In order to maximize the sensitivity of pulsar timing arrays to a stochastic
gravitational wave background, we present computational techniques to optimize
observing schedules. The techniques are applicable to both single and
multi-telescope experiments. The observing schedule is optimized for each
telescope by adjusting the observing time allocated to each pulsar while keeping
the total amount of observing time constant. The optimized schedule depends on
the timing noise characteristics of each individual pulsar as well as the 
performance of instrumentation.  Several examples are given
to illustrate the effects of different types of noise. A method to select the
most suitable pulsars to be included in a pulsar timing array project is
also presented. \end{abstract}


\section{Introduction}

Millisecond pulsars (MSPs) are stable celestial clocks, so that the
timing residuals, the differences between the observed and the predicted
time of arrival (TOA) of their pulses, are usually minute compared to
the total length of the data span. A stochastic gravitational wave (GW)
background leaves angular dependent correlations in the timing residuals
of widely separated pulsars (for general relativity see \citealt{HD83},
for alternative gravity theories see \citealt{ LJR08, LJR10}), i.e. the
correlation coefficient between timing residuals of a pulsar pair is
a function of the angular distance between the two pulsars. Such a
spatial correlation in pulsar timing signals makes it possible  to
directly detect GW using pulsar timing arrays (PTAs; \citealt{HD83,
FB90}). Previous analyses \citep{JHLM05} have calculated PTA sensitivity
to a stochastic GW background generated by super massive blackhole (SMBH)
binaries at cosmological distances \citep{JB03, SHMV04}. They have shown
that a positive detection of the GW background is feasible, if one uses
state of the art pulsar timing technologies. Such encouraging results
triggered consequent observational efforts.

At present, several groups are trying to detect GWs using PTAs:
i) the European Pulsar Timing Array (EPTA; \citealt{EPTA06, EPTA10,
FHB10, VLJ11}) with a sub-project, the Large European Array for
Pulsars (LEAP, \citealt{KS10, FHB10}), combining data from the Lovell
telescope, the Westerbork Synthesis Radio Telescope, the Effelsberg
100-m Radio Telescope, the Nan\c{c}ay Decimetric Radio Telescope, and
the Sardinia Radio Telescope\footnote{The Sardinia Radio telescope is
in the commissioning phase at the time of writing this this paper.},
ii) the Parkes Pulsar Timing Array (PPTA; \citealt{Man08,Hob09,
PPTA10}) using observations with the Parkes radio telescope augmented
by public domain observations from the Arecibo Observatory, iii)
the North-American Nanohertz Observatory for Gravitational waves
(NANOGrav, \citealt{Jenet09}) using data from the Green Bank Telescope
and the Arecibo Observatory, iv) Kalyazin Radio Astronomical Observatory
timing \citep{Rodin01}. Besides these on-going projects, international
cooperative efforts, e.g.~the International Pulsar Timing Array (IPTA,
\citealt{Ipta10}) or future telescopes with better sensitivity,
e.g.~the Five-hundred-meter Aperture Spherical Radio Telescope (FAST,
\citealt{NWZZJG04, SLKMSJN09}) and the Square Kilometre Array (SKA,
\citealt{KS10, SKSL09}), are planned to join the PTA projects to increase
the chances of detecting GWs.

Operational questions arise naturally from such PTA campaigns,
e.g.~how should the observing schedule be arranged to maximize our
opportunity to detect the GW signal? How much will we benefit from such
optimization? In this paper, we try to answer these questions. The
paper is organized as follows: In \SEC{sec:decs}, we extend the
formalism of \cite{JHLM05} to calculate the GW detection significance
as a function of observing schedules, i.e.\, the telescope time
allocation to each pulsar. Then we describe the technique to maximize
the GW detection significance in \SEC{sec:optintro}. Frameworks of the
optimization problem are described in \SEC{sec:optbak}, and the algorithm to
optimize a single and multiple telescope array are given in
\SEC{sec:optsig} and \SEC{sec:optmul} respectively. The results are presented in
\SEC{sec:res} and we discuss related issues in \SEC{sec:con}.

\section{Analytical Calculation For GW Detection Significance}
\label{sec:decs}

In this section, we calculate the statistical significance $S$ for
detecting the stochastic GW background using PTAs. We consider TOAs
from multiple pulsars, where each set may be collected from different
telescopes or data acquisition systems. To detect the GW background,
one correlates the TOAs between pulsar pairs and checks if the GW-induced
correlation is significant. \cite{JHLM05} have calculated the GW detection
significance for the case, where the noise in TOAs is of a white spectra
with equal root-mean-square (RMS) level for all pulsars. To investigate
the optimal observing schedule, we have to generalize the calculation,
such that we can explicitly check the dependence of the GW detection
significance on the noise properties of each individual pulsar.

Under the influence of a stochastic gravitational wave background, the
pulsar timing residual $R$ from a standard pulsar timing pipeline
contains two components, the GW-induced signal $\signalv$ and noise
from other contributions $\noisev$. In this section, we determine the
statistical properties of $\signalv$ and $\noisev$ first, and
then calculate the GW detection significance.

\subsection{Statistics for GW-induced pulsar timing signal}
\label{sec:psr}

The spectrum of the stochastic GW background is usually assumed to be
a power-law, in which the characteristic strain ($h_{\rm c}$) of the
GW background is $h_{\rm c}=A_{0} (f/f_{0})^{\alpha}$. Here, $A_{0}$
is the dimensionless amplitude for the background at $f_{0}=1\,
\textrm{yr}^{-1}$, and $\alpha$ is the spectral index. Under the
influence of such a GW background, the power spectrum $S_{\rm s}(f)$
of the GW-induced pulsar timing residual $\signalv$ is \citep{JHLM05}
\begin{equation}
S_{\rm s}(f)= \frac{A_{0}^2 f^{2\alpha-3}}{12 \pi^2
f_{0}^{2\alpha}}\,.
\label{eq:powsgw}
\end{equation}

GWs perturb the space-time metric at the Earth. This introduces a
correlation in the timing signal of two pulsars. The
correlation coefficient $H(\theta)$ between the GW-induced signals of
two pulsars with an angular separation of $\theta$ is called the
Hellings and Downs function \citep{HD83} given as 
\begin{equation}
H(\theta)=\left\{\begin{array}{l}
\frac{3+\cos\theta}{8}-\frac{3(\cos\theta-1)}{2} \ln
 \left[\sin\left(\frac{\theta}{2}\right)\right] \textrm{, if
 }\theta\neq 0\,, \\ 1 \textrm{, if }\theta=0\,.
 \end{array}\right. 
\label{eq:hdfun}
\end{equation}
The spectral properties, \EQ{eq:powsgw}, together with the
spatial correlation, \EQ{eq:hdfun}, fully characterize the
statistical properties of the GW-induced signals.

For an isotropic GW background, the correlations between the
GW-induced signals are 
\begin{equation} \langle \A s_{k}\B
 s_{k'}\rangle=\sigma_{\rm g}^2 H(\A \B \theta)
 \gamma_{kk'}\,. \label{eq:corgws}
\end{equation} 
Here, we follow the notation that the subscript on the right is the
index of sampling and the superscript on the left is the index for the
pulsar. For example, we denote the $k$-th measurement of a timing
residual of the $i$-th pulsar as $\A R_{k}$, the GW-induced signal as
$\A s_{k}$ and other noise contributions as $\A n_{k}$. $\sigma_{\rm
 g}$ is the RMS level for the GW-induced signal,
$\A \B \theta$ is the angular distance between the $i$-th and $j$-th
pulsar, and $\gamma_{kk'}$ is the temporal correlation coefficient
between the $k$-th and $k'$-th sampling. $\sigma_{\rm g}$ and
$\gamma_{kk'}$ are numerically calculated from the GW spectrum as
shown in \APP{sec:appcor}.

\subsection{Statistics of noise components from other contributions}

A purely theoretical modeling of the noise part $\noisev$ is complex,
because it \citep{FB90, SC10, CS10} depends on the properties of
each individual pulsar, the instrumentation, and radio frequency
interference. We therefore model the noise phenomenologically using
observational knowledge. In this paper, the noise part of a pulsar
timing residual is modeled as a superposition of a white noise component
and a red noise component, where the white noise is designated to the
measurement uncertainty on the TOA due to radiometer noise and pulse
jitter noise \citep{LKL12}. Timing residuals of several millisecond
pulsars show clear evidence of temporally correlated noise, although
its origin is not yet clear \citep{VBC09}. The red noise components
are used to empirically model such effects. We further assume that
the noise components are not correlated between any two different
pulsars\footnote{The correlated noise such as clock
 noise is discussed in \SEC{sec:con} and \APP{sec:cnm}}.

For each pulsar, three parameters are used to characterize the noise
spectrum. These parameters are the RMS level of the white noise $\A \sigma_{\rm 
w}$, the RMS level for red noise $\A
\sigma_{\rm r}$, and the spectral index for the red noise $\A \beta$.
The white noise spectrum is \begin{equation} \A S_{\rm
 w}(f)=\frac{\A\sigma_{\rm w}}{\widetilde{\sigma_{\rm w}}}\,,
 \label{eq:whitespec}
\end{equation} 
and the red noise spectrum is 
\begin{equation} \A S_{\rm
 r}(f)=\frac{\A\sigma_{\rm r} f^{\A\beta}}{ \widetilde{\sigma_{\rm
				r}}}\,,
 \label{eq:redspec} 
\end{equation} 
where the $\widetilde{\sigma_{\rm w}}$ and $\widetilde{\sigma_{\rm
 r}}$ are for normalization. From the spectrum, one can derive the
correlation between noise components
\begin{equation} \langle \A n_{k} \B n_{k'}\rangle=\left(\A \sigma_{\rm
 w}^2\delta_{kk'} +\A \sigma_{\rm r}^2 g_{kk'}\right)\delta_{ij}\,.
 \label{eq:cornoi} 
\end{equation} 
Following our conventions, $\A n_{k}$, is the noise for the $k$-th
sampling of the $i$-th pulsar. The $\delta_{ij}$ is the `Kronecker
delta' symbol, i.e. $\delta_{ij}=1$, if $i=j$, otherwise
$\delta_{ij}=0$. The parameters $\widetilde{\sigma_{\rm
 w}}$, $\widetilde{\sigma_{\rm r}}$, and correlation coefficients
$g_{kk'}$ are calculated using a numerical simulation shown in
\APP{sec:appcor}.

\subsection{GW background detection significance}
\label{sec:gwdecs}

Following \cite{JHLM05}, we calculate the cross power $\A \B c$
between timing residuals and then compare it with the predicted
correlation coefficient $H(\theta)$ to check whether the GW signal is
significant. The cross power $\A \B c$ is
\begin{equation}
 \A\B c=\frac{1}{m} \sumtim \Ra_{k} \Rb_k, \label{eq:defcorr}
\end{equation}
where $\Ra_{k}, \Rb_{k}$ are the timing residuals of the $i$-th and
the $j$-th pulsars for the $k$-th timing session, $m$ is the number of
data points for a given pulsar. It is assumed that the data from
different pulsars overlap with each other and the number of data
points is identical for all the pulsars, similar to the discussion in
\cite{YCH11}. These assumptions are good approximations in calculating the GW 
detection sensitivity. In
the case where the TOAs are mis-aligned or the number of data
points is not identical, one can combine data points so that the
number of data points are identical. This operation retains most of
the GW detection sensitivity due to: i) The RMS error reduces when
data is combined, and such a reduction in RMS error compensates the
reduction in the number of data points. ii) The spectrum of the GW
induced timing signal is steep, thus most of the GW induced signal is
in the low frequency components, which are preserved during the
combining operation.

The comparison between the $\A\B c$ and the Hellings-Downs function is
carried out by doing another correlation, which gives the GW detection
significance $S$ as
\begin{equation}
 S=\sqrt{M} \frac{\sum_{\ijp} (\A \B c-\overline{c}) (H(\A \B
 \theta)-\overline{H}) }{\sqrt{\sum_{ \ijp}(\A \B c-\overline{c})^2
 \sum_{\ijp} (H(\A \B \theta)-\overline{H})^2}},
\label{eq:decs}
\end{equation}
where the summation $\sum_{\ijp}$ sums over all independent pulsar
pairs except the case where $i=j$, i.e.
\begin{equation}
 \sum_{\ijp}\equiv\sum_{i=1}^{N}\sum_{j=1}^{i-1}\,,\end{equation}
and \begin{eqnarray}\overline{ c}&=&\frac{1}{M}\sum_{\ijp} \A \B c\,,
 \\ \overline{H}&=&\frac{1}{M}\sum_{\ijp} H(\A \B \theta)\,.
\end{eqnarray}
Given $N$ pulsars, the sum $M=\sum_{\ijp} 1=N(N-1)/2$ is the number of
independent pulsar pairs. To evaluate the quality of the detector, we
need the expectation for the detection significance $\langle S
\rangle$, which is
\begin{equation} \langle S \rangle \simeq
 \frac{\sum_{ \ijp} \langle (\A \B c-\overline{\langle c\rangle})
 (H(\A \B \theta)-\overline{H}) \rangle }{\sqrt{M}\Sigma_c
 \Sigma_H} \label{eq:exps}, \end{equation} 
where
\begin{eqnarray} \Sigma_{\rm
 H}&=&\sqrt{\frac{1}{M}\sum_{ \ijp} [H(\A \B
 \theta)-\overline{H}]^2}\,, \\ \Sigma_{c}&=&\sqrt{\left
 \langle { \frac{1}{M}\sum_{ \ijp} [c(\A \B
 \theta)-\overline{\langle c\rangle}]^2} \right\rangle}\,, 
\end{eqnarray}
and the $\langle \cdot \rangle$ denotes the ensemble
average. As we show in Appendix~\ref{sec:appS}, the
expected GW detection significance is 
\begin{equation}
 \langle S\rangle \simeq \sqrt{M} \left[{1+\frac{
  \sum_{\ijp}\left( \A\B A+ \A\B B\right) }{M
  \Sigma_{\rm H}^2 }}\right]^{-\frac{1}{2}}\,,
  \label{eq:decse}
\end{equation}
where 
\begin{eqnarray} \A\B
  A&=&\frac{1}{m^2}\sum_{kk'}\left[\left(1+H(\A \B
  \theta)\right)\gamma_{kk'}^2 \nonumber
  \right.\\ &&\left.+\left(\A \eta_{ {\rm r}, kk'} + \B
  \eta_{ {\rm r}, kk'}\right){\gamma_{kk'}}+\A\eta_{ {\rm
   r}, kk'} \B\eta_{ {\rm r}, kk'}\right] \,,\\ \A\B
  B&=& \frac{1}{m}\left(\A \eta_{ \rm w} +\B \eta_{\rm
  w}+\A\eta_{\rm w}\B\eta_{\rm w} +\A\eta_{\rm r}
  \B\eta_{\rm w} +\B\eta_{\rm r}\A\eta_{\rm
  w}\right)\,, \label{eq:AB}
\end{eqnarray}
and $\eta$ denotes the ratio between the power of noise components and
the power of the GW-induced signal, which are 
\begin{eqnarray} \A
 \eta_{ {\rm r}, kk'}&=& \frac{\A \sigma_{\rm r}^2 \A
 g_{kk'}}{\sigma_{\rm g}^2}\, \\ \A \eta_{\rm r}&=& \frac{\A
 \sigma_{\rm r}^2}{\sigma_{\rm g}^2}\, \\ \A \eta_{\rm w}&=&
 \frac{\A \sigma_{\rm w}^2}{\sigma_{\rm g}^2}\,.
\end{eqnarray} 
If the noise level $\A \sigma_{\rm n}$ is identical for all pulsars
and there is no red noise component ($\A \sigma_{\rm w}=\B \sigma_{\rm
 w}$ and $\A \sigma_{\rm r}=0$ for all the $i,j=1\ldots N$),
\EQ{eq:decse} reduces to the result found by \cite{JHLM05} and
\cite{VBC09}.

The $\A\B A$ terms are independent of telescope integration time,
because they only contain the RMS level of the GW-induced signal and
the red noise, both of which are independent of telescope integration
time. The observing schedule, the plan of allocating telescope time
to each pulsar, only changes terms of $\A\B B$, which depend on the
ratio between signal and noise amplitude. By optimizing the
observing schedule, we can reduce the summation of $\A\B B$, such
that the GW detection significance is maximized. We present the
technique to maximize the GW detection significance in the next
section.

\section{Optimization Of The GW Detection Significance Under Observing
Constraints}
\label{sec:optintro}

Pulsar timing observations are usually conducted as a series of
successive observing sessions. In each session, multiple pulsars are
observed using either one or multiple telescopes. There are basically
two ways to use multiple telescopes. The simple way is to use each
telescope independently, combine the TOA data from each telescope,
remove the time jumps between each data set, and form a single TOA
data set \citep{JSKN08}. The other way, as in the LEAP project,
is to use telescopes simultaneously to form a phased array and then
calculate the TOAs from the phased-array data. Due to such different
methods of using multiple telescopes, the optimization techniques differ
from one another. We answer the following questions in this section: i)
What are the variables to optimize? ii) What are the constraints on
the optimization? iii) How do we perform the optimization?

\subsection{The objective, variables, and constraints of the optimization}
\label{sec:optbak}

Given a fixed amount of telescope time, we can adjust the amount of
observing time allocated to each pulsar. Increasing the observing time
for one pulsar will reduce its timing measurement error, but will
increase the timing error for other pulsars. Naturally, for the
purpose of detecting GWs, the optimization objective is to maximize
the expected GW detection significance $\langle S \rangle$, while the
constraint is the total amount of telescope time for the project.

Generally, for a timing project using $\ntel$ telescopes to observe
$N$ pulsars, we need $2\ntel+3N$ input parameters to characterize the
whole timing project. $2\ntel$ parameters are used to characterize
$\ntel$ telescopes, where each telescope is quantified by the
\emph{gain} ($\gain$) and the \emph{total available telescope time}
($\tau$). $3N$ parameters are used to characterize the timing behavior
of $N$ pulsars. The red noise level
$\sigma_{\rm r}$ and spectral index $\beta$ are assumed to be pulsar
intrinsic. The observed white noise RMS levels $\sigma_{\rm w}$
depend on telescope gain, telescope time allocation to the pulsar and
other parameters intrinsic to the pulsar (e.g.~flux, pulse
width, and so on).

For single telescope cases, we can encapsulate the dependence of
the white noise level on the schedule into the normalized white
noise level $\sigma_{\rm 0}$ and pulse jitter noise level $\sigma_{\rm J}$
\citep{FB90, SC10, CS10} 
\begin{equation}
 \A \sigma_{\rm w}=\left(\A \sigma_{0}^2 \gain^{-2} +\A \sigma_{\rm
 J}^2\right)^{1/2}\left(\frac{\A \tau}{1 \textrm{hr}}\right)^{-1/2}
 , \label{eq:scale}
\end{equation} 
where $\A \sigma_{\rm w}$ is the measured RMS level for the white
noise component of the $i$-th pulsar, $\A \tau$ is the telescope time
being used for the $i$-th pulsar per observing session, and $\gain$ is
the telescope gain. The normalized noise level $\A
\sigma_{0}$ and the $\A \sigma_{\rm J}$ are used to characterize the
radiometer noise and the pulse jitter noise of the pulsar. On the one
hand, if there is no pulse jitter noise, $\sigma_{\rm 0}$ will be the
observed RMS level of the white noise for a 1 hour observation using a
telescope with unit gain $\gain=1$\footnote{In this paper, we use a
 fiducial unit for the gain, thus one can take any telescope as unit
 1 and scale other telescopes accordingly. }. On the other hand, if
we have a telescope with infinite gain, the pulsar timing accuracy
will be limited by the pulse jitter, and $\sigma_{\rm J}$ will be the
observed RMS level of white noise for a 1 hour observation. If phased
array observations are performed using multiple telescopes (as done in
the LEAP project), the situation is identical to the case of a single
telescope, and we use the effective gain for the array to determine
the RMS level of noise.

For multiple incoherent telescopes, the gain of $\ntel$ telescopes can
be summarized by a vector $\gain_{\nu}$, where $\nu=1\ldots\ntel$ and
the $\nu$-th component $\gain_{\nu}$ is the gain for the $\nu$-th
telescope. In a similar fashion, the available telescope time of each
telescope is summarized by the vector $\tau_{\nu}$. The definition of
the observing schedule becomes more complex, since we need to
specify the observation time for each pulsar using each
telescope. Furthermore, the schedule should also include information on the 
telescope availability, e.g.~certain pulsars may not be
visible to some telescopes due to geographical reasons. In this paper,
we use the \emph{resource allocation matrix} ${\rm \bf O}$ to describe
the telescope availability, where the $i$-th row and $\nu$-th column
element $\A O_{\nu}$ indicates whether we use the $\nu$-th telescope
to observe the $i$-th pulsar, i.e.\,$\A O_{\nu}=1$, if the $\nu$-th
telescope observes the $i$-th pulsar, and $\A O_{\nu}=0$
otherwise. The telescope time allocation is described by another
matrix, the \emph{schedule matrix} ${\rm \bf P}$, where the $i$-th row
$\nu$-th column element $\A P_{ \nu}$ is the time allocated for the
$\nu$-th telescope observing the $i$-th pulsar.

With the schedule matrix ${\rm \bf P}$, the equivalent RMS level of the
white noise in the combined data is
\begin{equation}
 \A \sigma_{\rm w}=\left[\A \sigma_{0}^2 \left(\sum_{\nu=1}^{\ntel}
 \gain_{\nu}^2\A P_{\nu} \A O_{\nu}
 \right)^{-1}+\A \sigma_{\rm J}^2 \left(\sum_{\nu=1}^{\ntel}\A P_{\nu} \A
 O_{\nu}
 \right)^{-1}\right]^{1/2}\,.
 \label{eq:incadd}
\end{equation}
The main reason to introduce the resource allocation matrix $\A
O_{\nu}$ is to take care of the complexity of telescope
availability, and the telescope time can be treated in an identical
way independent of the availability.

For a single telescope or a phased array, the optimization constraint
is \begin{equation} \tau=\sum_{i=1}^{N} \A \tau\,,
 \label{eq:cons1}
\end{equation}
i.e. the total telescope time is pre-fixed to be $\tau$. For multiple
incoherent telescopes, the above constraint is applied to each
telescope individually, i.e.\,the constraints specify the available
observation time $\tau_{\nu}$ for each telescope, which
gives 
\begin{equation} \tau_{\nu}=\sum_{i=1}^{N}\A P_{\nu} \A
 O_{\nu}\,.
 \label{eq:cons3}
\end{equation}

With the observing schedule (i.e.\,the vector $\A\tau$ for a single
telescope or the matrix $\A O_{\nu}$ and $\A P_{\nu}$ for incoherent
telescopes), one can use \EQ{eq:scale} and ~(\ref{eq:incadd}) to
determine the white noise level, then determine the GW detection
significance as explained in \SEC{sec:gwdecs}. The optimization of
the observing schedule means that we choose an appropriate $\A \tau$
or $\A P_{\nu}$ such that the expected GW-detection significance
$\langle S \rangle$ is maximized under the constraint of \EQ{eq:cons1}
or (\ref{eq:cons3}).

From \EQ{eq:decse}, the maximization of $\langle S \rangle$ is
equivalent to the minimization of the term $\sum_{\ijp} (\A\B A+\A\B
B)$. Because the terms of $\A\B A$ are independent of the telescope
time, the maximization of $\langle S \rangle$ by adjusting the
observing schedule is thus equivalent to minimizing the objective
function $\Lag$ defined as
\begin{equation}
\Lag=\sum_{\ijp} \A\B B\,.
 \label{eq:lag}
\end{equation}

\subsection{Optimizing a single telescope}
\label{sec:optsig}

Our technique to optimize the single telescope schedule takes two
steps. The first step is to convert the constrained optimization
problem to a constraint-free version, and the second step is to solve
the constraint-free optimization. For comparison, we have also
developed an alternative semi-analytical iterative technique in
Appendix~\ref{sec:appopt}.

To remove the constraints, we transform variables $\A \tau$ to a new
set of variables $\theta_\mu$, where $\mu$ is $1\ldots N-1$, and the
transformation is
\begin{equation}
 \left(\begin{array}{c}
 \theta_1 \\
 \theta_2 \\
 \theta_3\\
 \ldots \\
 \theta_{N-2} \\
 \theta_{N-1}
\end{array}\right)=\left(\begin{array}{c}
 \arccos \left(\frac{\sqrt{\,^{1}\negthinspace \tau }}{\sqrt{\tau }}\right) \\
 \arccos \left(\frac{\sqrt{\,^{2}\negthinspace \tau }}{\sqrt{\tau}
 \sin\theta_1}\right) \\
 \arccos \left(\frac{\sqrt{\,^{3}\negthinspace \tau }}{\sqrt{
 \tau}\sin\theta_1\sin\theta_2 }\right)\\
 \ldots \\
 \arccos \left(\frac{\sqrt{\,^{N-2}\negthinspace \tau }}{\sqrt{\tau}
 \prod_{i=1}^{N-3}\sin\theta_i }\right) \\
 \arccos \left(\frac{\sqrt{\,^{N-1}\negthinspace \tau }}{\sqrt{\tau}
 \prod_{i=1}^{N-2}\sin\theta_i }\right)
 \end{array}\right)\,,
 \label{eq:transforma}
\end{equation}
of which the inverse transformation is \begin{equation}
 \left(\begin{array}{c}
 {\,^{1}\negthinspace \tau } \\
 {\,^{2}\negthinspace \tau } \\
 {\,^{3}\negthinspace \tau } \\
 \ldots \\
 {\,^{N-1} \tau } \\
 {\,^{N} \tau }
 \end{array}\right)=
 \tau \left(\begin{array}{c}
 \cos^2 \theta_1 \\
 \sin^2 \theta_1 \cos^2 \theta_2 \\
 \sin^2 \theta_1 \sin^2 \theta_2 \cos^2 \theta_3\\
 \ldots \\
 \prod_{\mu=1}^{N-2}\sin^2\theta_\mu \cos^2 \theta_{N-1} \\
 \prod_{\mu=1}^{N-1}\sin^2\theta_\mu\end{array}\right)\,.
 \label{eq:transform}
\end{equation}
Here it is implicitly assumed that $\A \tau \ge 0$, the $\prod_i$ and
$\prod_\mu$ are the serial products using the index $i$ and $\mu$
respectively. Equations (\ref{eq:transforma}) and (\ref{eq:transform})
are, in fact, the transformation between a $N$-dimensional Cartesian
coordinate system and its corresponding hyperspherical coordinate
system. The constraint in the new spherical coordinate system
corresponds to fixing the radius of the hypersphere (fixing $\tau$),
and all angular coordinates $\theta_{\mu}$ are free variables for
which one can specify any value for $\theta_{\mu}$ without breaking
the constraint, i.e.\, after the above coordinate transformation, the
objective $\Lag$ becomes a function of a new variable $\theta_\mu$,
the constraint \EQ{eq:cons1} is automatically satisfied for any
$\theta_{\mu}$. We can then find the minimum of $\Lag$ using
numerical methods for constraint-free problems. In this paper, the
downhill simplex method \citep{NM65} is adopted, which has, usually,
better global converging behavior than other methods \citep{KWJ94}.
Once the optimal $\theta_\mu$ are found, we transform them back to $\A
\tau$ using \EQ{eq:transform}, which yields the optimal single
telescope schedule. In practical situations, one also needs to add
telescope slewing time, observing time for calibration sources and
other necessary auxiliary time on top of this telescope time schedule
to get the final schedule.

\subsection{Optimizing multiple incoherent telescopes}
\label{sec:optmul}

Similarly to section~\ref{sec:optsig}, the optimal observing schedule for 
multiple
incoherent telescopes is calculated by minimizing $\Lag$, although the
constraints are slightly more complex here. In this section, we present the
optimizing technique and discuss later the relation between the optimization of
multiple telescopes and single telescope optimization.

From the multiple telescope constraint \EQ{eq:cons3}, we can see that
the constraints are applied to each telescope individually. Thus, the
generalization of the method presented in the previous section is
straightforward by applying the transformation \EQ{eq:transforma} to
each telescope separately. Take telescope 1 as an example. The first
column of matrix $\rm \bf P$, the $\,^{1} \negthinspace P_{\nu}$, is
the observing schedule for the first telescope. We can transform
those components of $\,^{1} P_{\nu}$ indicated by
$\,^{1} \negthinspace O_{\nu}=1$ using \EQ{eq:transforma} to remove
the constraint of the first telescope. Similarly, by applying the
transformation to the other columns of matrix $\rm \bf P$
successively, one can remove all the constraints. With the new
constraint-free variables, we use the down-hill simplex method to find
the optimization. We then transform back to $\A P_{\nu}$, which is the
optimization schedule.

In the optimization algorithm, we treat the resource allocation
matrix $\rm \bf O$ as input knowledge. A question naturally arises
as to whether one can find a better observing schedule for the same
telescopes with the same amounts of telescope time but with a different
$\rm \bf O$, i.e.\,whether one can add pulsars to or remove pulsars
from schedules of certain telescopes to increase the detection
significance? The configuration with all $\A O_{\nu}=1$ allows one to
use any telescope to observe any pulsar, i.e. allows one to adjust the
schedule with the maximal degrees of freedom. In this way, the
optimization schedule of configuration ($\A O_{\nu}=1$) leads to the
highest GW detection significance. If any schedule has the same
detection significance as the optimal schedule with all $\A
O_{\nu}=1$, we call such schedule `global optimal'. To determine whether
the global optimization is achievable, we first investigate the case
when the pulse jitter noise can be ignored, we then discuss situations where the 
pulse jitter noise becomes important.

When the pulse jitter noise is neglected, there is a close relation
between the single telescope optimization and the multiple telescope
optimization. In fact, under certain conditions, the optimization for
multiple telescopes is equivalent to the single telescope
optimization. To see this, we replace the variables $\A P_{\nu}$ and
$\A O_{\nu}$ in \EQ{eq:incadd} by \emph{effective telescope time} $\A
\tau_{\rm e}$ as follows 
\begin{equation} \A\tau_{\rm
 e}=\sum_{\nu=1}^{N_{\rm tel}} \gain_{\nu}^2 \A P_{\nu} \A
 O_{\nu}\,.
 \label{eq:taueff}
\end{equation}
After ignoring the $\sigma_{\rm J}$, \EQ{eq:incadd}
becomes 
\begin{equation} \A \sigma_{\rm w}=\A \sigma_{0} \A \tau_{\rm
 e}^{-1/2} \,,
 \label{eq:effopta}
\end{equation}
and the constrains \EQ{eq:cons3} reduces to a single
constraint 
\begin{equation} \sum_{i=1}^{N} \A \tau_{\rm e} = \tau_{\rm
 e}\equiv \sum_{\nu=1}^{N_{\rm tel}} \tau_{\nu} \gain_{\nu}^2 \,.
 \label{eq:effopt}
\end{equation}
By comparing equations (\ref{eq:effopta}) and (\ref{eq:effopt}) with
\EQ{eq:cons1} and (\ref{eq:scale}), one can see that the optimization
for multiple telescopes is very similar to the single telescope
optimization. In fact, if there exists a unique solution of $\A
P_{\nu}$ to \EQ{eq:taueff}, which satisfies each individual constraint
of \EQ{eq:cons3}, the multiple telescope and single telescope
optimization are mathematically identical.

The differences between multiple telescope and single telescope
optimization lie in the difference between the constraints
\EQ{eq:cons3} and (\ref{eq:effopt}). For the single telescope case,
only one constraint (equation \ref{eq:cons1}) is involved. This is very
different from the case of multiple telescopes, where $N_{\rm tel}$
constraints (equation \ref{eq:cons3}) are present. The variable substitution in
\EQ{eq:taueff} combines all $N_{\rm tel}$ constraints (equation \ref{eq:cons3})
and forms a single constraint (equation \ref{eq:effopt}), where the constraint
of each individual telescope is ignored. Whether global optimization
is achievable is, now, equivalent to whether one can find a solution
to $\A P_{\nu}$ for \EQ{eq:taueff}, while satisfying all individual
constraints (equation \ref{eq:cons3}).

Generally, when solving $\A P_{\nu}$ from \EQ{eq:taueff} and
(\ref{eq:cons3}), one meets three types of situations: a unique
solution, multiple solutions and no solution. For the case of multiple
solutions, there are multiple choices for the optimal schedule. All
these configurations are identical in the sense that they give the
same GW detection significance. The case of no solution can only arise
when constraints for some telescopes cannot be met, i.e.\,some of the
pulsars need more time than the telescopes can give, while some of the
pulsars have more telescope time than should be assigned. Take the
case with two telescopes and two pulsars as an example, where the gain
of the two telescopes are the same, two pulsars have identical
$\sigma_{0}$, and $\A O_{\nu}=\left[\begin{array}{c c}1& 1 \\0&
 1\end{array}\right]$, i.e.\,only the second telescope can observe
the second pulsar. Since the two pulsars have the same $\sigma_{0}$, the
total effective telescope time should then also be the same for the
optimal schedule (i.e.\,$\,^{1} \tau_{\rm e}= \,^{2} \tau_{\rm
 e}$). However if the first telescope does not have enough time, one
gets $\,^{1} \tau_{\rm e}< \,^{2} \tau_{\rm e}$ and the global optimal
schedule is not achievable for such configurations.

The case for which there no solution to \EQ{eq:taueff}, is due to an improper
choice of telescopes. Most of the time the matrix $\rm \bf O$ is
determined by the sky coverage of the telescopes. Thus, if no solution
can be found, one needs to seek telescopes with the appropriate sky
coverage or extend the time for telescopes with the sky coverage. In
order to identify such a no-solution situation we show in
\APP{app:exist} that a solution to \EQ{eq:taueff} exists, if the
following conditions are satisfied 
\begin{eqnarray} \A \tau_{\rm e}\le
 \sum_{\nu=1}^{N_{\rm tel}} \gain^2 \tau_{\nu} \A O_{\nu}\,, \textrm{
 for any index of pulsar }i.
 \label{eq:existsol}
\end{eqnarray} 
One can identify which pulsar in \EQ{eq:existsol} fails. These
failures indicate the corresponding elements of resources allocation
matrix to be adjusted.

We now consider the situation where the pulse jitter noise becomes
important. Clearly, if we cannot ignore the pulse jitter noise, the
effective telescope time is no longer linearly dependent on the
telescope time $\A P_{\nu}$ as in \EQ{eq:taueff} and we do not have a
simple method to check if the global optimization is achieved. However
we can still set all the $\A O_{\nu}=1$, find the global optimal
strategy, and compare it with the optimal strategy for the input $\A
O_{\nu}$ to check if the global optimization is achieved.

\begin{figure*}
 \centering \includegraphics[totalheight=4.5in]{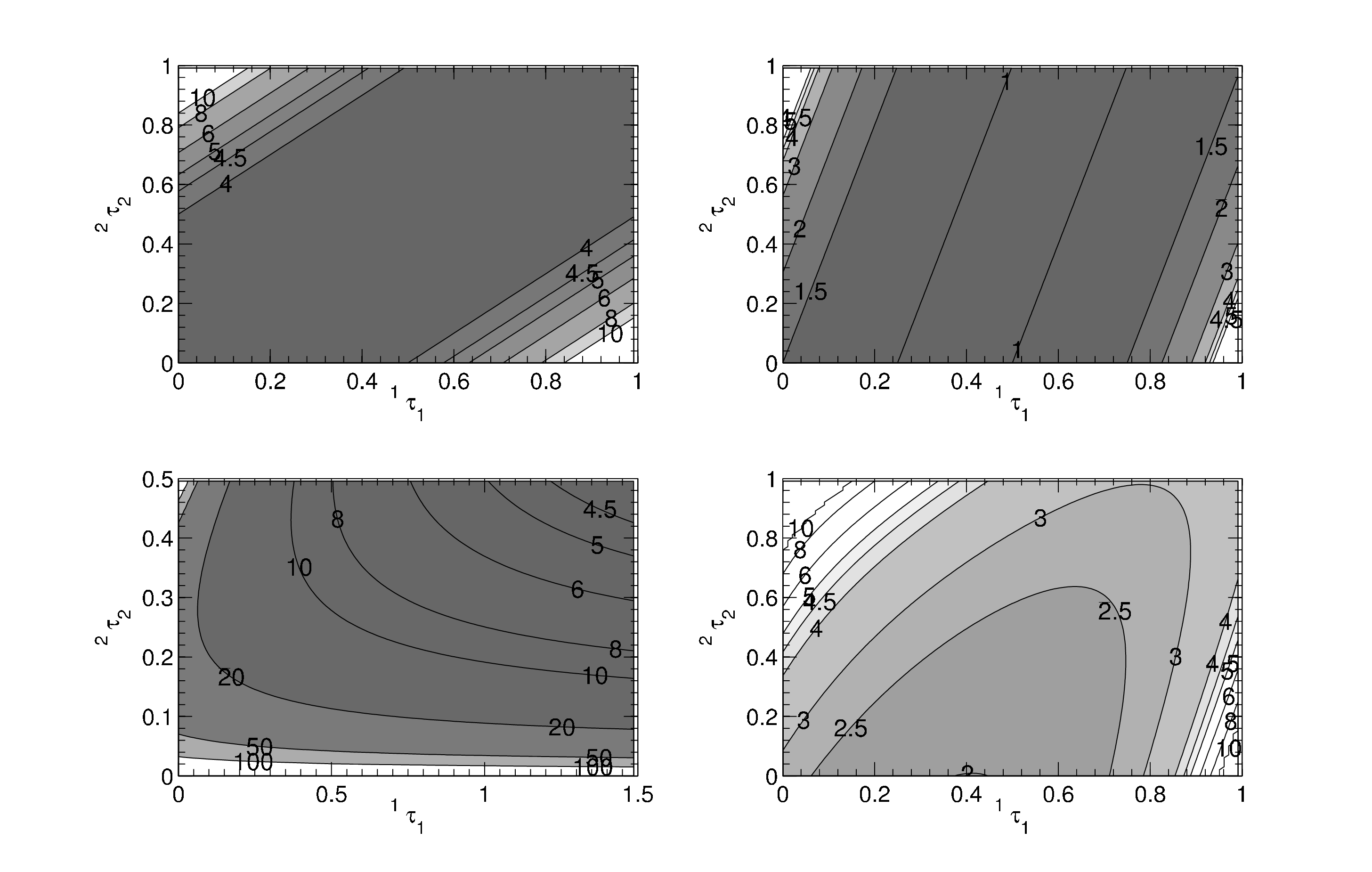}
\caption{Four examples to illustrate the optimization of multiple
 telescopes. Here, we plot $\Lag$, defined in \EQ{eq:lag}, as a
 function of telescope time. The smaller $\Lag$ is, the better the
 observing schedule. In all examples, two pulsars are observed with
 two telescopes. The telescope configurations and pulsar noise
 parameters for each case is given in \TAB{tab:ex4}. The x-axis and
 y-axis are for the $\,^{1}\tau_{1}$, i.e.\, the time for the 1st
 telescope observing the 1st pulsar, and $\,^{2}\tau_{2}$, i.e.\, the
 time for the 2nd telescope observing the 2nd pulsar,
 respectively. \label{fig:mulexam}}
\end{figure*}

In \FIG{fig:mulexam}, we give four examples for the optimization of
multiple telescopes. In these examples, two telescopes are used to
observe two pulsars, where the pulsar noise parameters and the
telescope parameters are specified in \TAB{tab:ex4}. These four
examples are given as follows. i) `Case A', two identical pulsars are
observed with two identical telescopes. The global optimal schedule is
achievable and one can exchange telescope time between the two
telescopes. As indicated in \FIG{fig:mulexam}, as long as we assign the same 
amount of effective
telescope time $\A \tau_{\rm e}$ for two pulsars, it is the optimal
observing schedule. ii) `Case B', telescope 1 has twice the gain
compared to telescope 2. Similar to `case A', the global optimal
schedule is achievable and one can exchange telescope time. But due to
the gain differences, the telescope time exchange should be weighted
by the gain, such that the same amount of effective telescope time is
assigned to the two identical pulsars. iii) `Case C', where telescope
1 has more time (1.5 hr) available compared to telescope 2 (1 hr) and
only telescope 2 can be used to observe the 2nd pulsar (as indicated
by the matrix $\A O_{\nu}$). For this case, the total telescope time is
the same as in `case A', but we do not achieve the same low level of
$\Lag$ as in `case A', i.e.\,global optimization is not
achievable. This is due to the constraint from matrix $\A O_{\nu}$,
which prevents us from reaching the global optimization as
discussed. iv) `Case D', where pulsar 1 is affected by pulse
jitter noise. In this case, one does not have the freedom to exchange
telescope time and optimization suggests that the low gain telescope
(telescope 2) should spend more time on the pulse jitter affected
pulsar (pulsar 1).

\begin{table*}
\begin{center}
\caption{The telescope configurations and pulsar noise parameters for
 each case given in \FIG{fig:mulexam}. $\sigma_{0}$ and $\sigma_{\rm
 J}$ take the unit of second, $\tau$ takes the unit of hour, and
 the gain $\gain$ take an arbitrary fiducial unit as explained in the
 main text. \label{tab:ex4} }
\begin{tabular}{l lll c lll}\hline\hline
 &case A: & & & & case B: & &\\ \\ Radiometer noise& $\,^{1}
 \sigma_{0}=10^{-7}$ & $\,^{2} \sigma_{0}=10^{-7}$ & & & $\,^{1}
 \sigma_{0}=10^{-7}$ & $\,^{2} \sigma_{0}=10^{-7}$ & \\ Jitter
 noise&$\,^{1} \sigma_{\rm J}=0$ & $\,^{2} \sigma_{\rm J}=0$ & & &
 $\,^{1} \sigma_{\rm J}=0$ & $\,^{2} \sigma_{\rm J}=0$ &
 \\ Observation time&$\tau_{1}=1$ & $\tau_{2}=1$ & & & $\tau_{1}=1$ &
 $\tau_{2}=1$ & \\ Telescope gain&$\gain_{1}=1$ & $\gain_{2}=1$ & & &
 $\gain_{1}=2$ & $\gain_{2}=1$ &\\ Resource allocation matrix&$\A
 O_{\nu}=\left[\begin{array}{c c}1& 1 \\ 1& 1\end{array}\right]$ & &
 & & $\A O_{\nu}=\left[\begin{array}{c c}1& 1 \\ 1&
 1\end{array}\right]$ & &\\ \hline &case C: & & & & case D: &
 &\\ \\ Radiometer noise&$\,^{1} \sigma_{0}=10^{-7}$ & $\,^{2}
 \sigma_{0}=10^{-7}$ & & & $\,^{1} \sigma_{0}=10^{-7}$ & $\,^{2}
 \sigma_{0}=10^{-7}$ & \\ Jitter noise &$\,^{1} \sigma_{\rm J}=0$ &
 $\,^{2} \sigma_{\rm J}=0$ & & & $\,^{1} \sigma_{\rm J}=10^{-7}$ &
 $\,^{2} \sigma_{\rm J}=0$ & \\ Observation time&$\tau_{1}=1.5$ &
 $\tau_{2}=0.5$ & & & $\tau_{1}=1$ & $\tau_{2}=1$ & \\ Telescope
 gain&$\gain_{1}=1$ & $\gain_{2}=1$ & & & $\gain_{1}=2$ &
 $\gain_{2}=1$ &\\ Resource allocation matrix&$\A
 O_{\nu}=\left[\begin{array}{c c}1& 1 \\ 0& 1\end{array}\right]$ & &
 & & $\A O_{\nu}=\left[\begin{array}{c c}1& 1 \\ 1&
 1\end{array}\right]$ & &\\ \hline \hline
\end{tabular}
\end{center}
\end{table*}

\section{Results}
\label{sec:res}

As examples, we use pulsar properties measured from data of the PPTA,
EPTA and NANOGrav to show the potential benefit of optimizing the
observing schedule. The parameters of the pulsars are given in
\TAB{tab:psrp}. 

\begin{table*}
\begin{center}
 \caption{Parameters for PPTA, EPTA and NANOGrav pulsars taken from
 Hobbs et al. (2010). We assume that all the white noise is due
 to the radiometer noise and pulse jitter noise can be ignored.
\label{tab:psrp}
}
\begin{tabular}{cccccccc}\hline\hline
 PSR J & P & $P_b$ & S1400 & Array & PPTA $\sigma_{0}$ & EPTA $\sigma_{0}$ &
 NANOGrav $\sigma_{0}$ \\
   & (ms) & (d) & (mJy) & & ($\mu$s)& ($\mu$s)& ($\mu$s) \\ \hline
J0030$+$0451 & 4.87 & - & 0.6 & EPTA, NANOGrav & - & 0.54 & 0.31 \\
J0218$+$4232 & 2.32 & 2.03 & 0.9 & NANOGrav & - & - & 4.81\\
J0437$-$4715 & 5.76 & 5.74 & 142.0 & PPTA & 0.03 & - & - \\
J0613$-$0200 & 3.06 & 1.20 & 1.4 & PPTA, EPTA, NANOGrav & 0.71 & 0.45 & 0.50\\
J0621$+$1002 & 28.85 & 8.32& 1.9& EPTA & - & 9.58 & - \\

J0711$-$6830 & 5.49 & - & 1.6 & PPTA & 1.32 & - & - \\
J0751$+$1807 & 3.48 & 0.3 & 3.2 & EPTA & - & 0.78 & - \\
J0900$-$3144 & 11.1 & 18.7 & 3.8 & EPTA & - & 1.55 & - \\
J1012$+$5307 & 5.26 & 0.60 & 3.0 & EPTA, NANOGrav & - & 0.32 & 0.61\\
J1022$+$1001 &16.45 & 7.81 & 3.0 & PPTA, EPTA & 0.37 & 0.48 & - \\
J1024$-$0719 & 5.16 & - & 0.7 & PPTA, EPTA & 0.43 & 0.25 & -\\
J1045$-$4509 & 7.47 & 4.08 & 3.0 & PPTA & 2.68 & - & - \\
J1455$-$3330 & 7.99 & 76.17 & 1.2 & EPTA, NANOGrav & - & 3.83 & 1.60\\
J1600$-$3053 & 3.60 & 14.35 & 3.2 & EPTA, PPTA & 0.32 & 0.23 & - \\
J1603$-$7202 & 14.84 & 6.31 & 3.0 & PPTA & 0.70 & - & - \\

J1640$+$2224 & 3.16 & 175.46 & 2.0 & EPTA, NANOGrav & - & 0.45 & 0.19\\
J1643$-$1224 & 4.62 & 147.02 & 4.8 & PPTA, EPTA, NANOGrav & 0.57 & 0.56 & 0.53\\
J1713$+$0747 & 4.57 & 67.83 & 8.0 & PPTA, EPTA, NANOGrav & 0.15 & 0.07 & 0.04 \\
J1730$-$2304 & 8.12 & - & 4.0 & PPTA, EPTA & 0.83 & 1.01 & - \\
J1732$-$5049 & 5.31 & 5.26 & - & PPTA & 1.74 & - & - \\

J1738$+$0333 & 5.85 & 0.35 & - & NANOGrav & - & - & 0.24 \\
J1741$+$1351 & 3.75 & 16.34 & - & NANOGrav & - & - & 0.19 \\
J1744$-$1134 & 4.08 & - & 3.0 & PPTA, EPTA, NANOGrav & 0.21 & 0.14 & 0.14\\
J1751$-$2857 & 3.91 & 110.7 & 0.06 & EPTA & - & 0.90 & - \\
J1824$-$2452 & 3.05 & - & 0.2 & PPTA, EPTA & 0.39 & 0.24 & -\\
J1853$+$1303 & 4.09 & 115.65 & 0.4 & NANOGrav &- & -& 0.17 \\
J1857$+$0943 & 5.37 & 12.33 & 5.0 & PPTA, EPTA, NANOGrav & 0.82 &
0.44 & 0.25 \\
J1909$-$3744 & 2.95 & 1.53 & 3.0 & PPTA, EPTA, NANOGrav & 0.19 &
0.04 & 0.15\\
J1910$+$1256 & 4.98 & 58.47 & 0.5& EPTA, NANOGrav & - & 0.99 & 0.17 \\
J1918$-$0642 & 7.65 & 10.91 & - & EPTA, NANOGrav & - & 0.87 & 1.08\\

J1939$+$2134 & 1.56 & - &10.0 & PPTA, EPTA, NANOGrav & 0.11 &
0.02 & 0.03 \\
J1955$+$2908 & 6.13 & 117.35 & 1.1 & NANOGrav & - & - & 0.18 \\
J2019$+$2425 & 3.94 & 76.51 & - & NANOGrav & - & - & 0.66 \\
J2124$-$3358 & 4.93 & - & 1.6 & PPTA & 1.52 & - & -\\
J2129$-$5721 & 3.73 & 6.63 & 1.4 & PPTA & 0.87 & - & -\\
J2145$-$0750 & 16.05 & 6.84 & 8.0 & PPTA, EPTA, NANOGrav & 0.86 &
0.40 & 1.37\\
J2317$+$1439 & 3.44 & 2.46 & 4.0 & NANOGrav &- & 0.81 & 0.25\\
\hline \hline
\end{tabular}
\end{center}
\end{table*}

\begin{figure*}
 \centering \includegraphics[totalheight=4in]{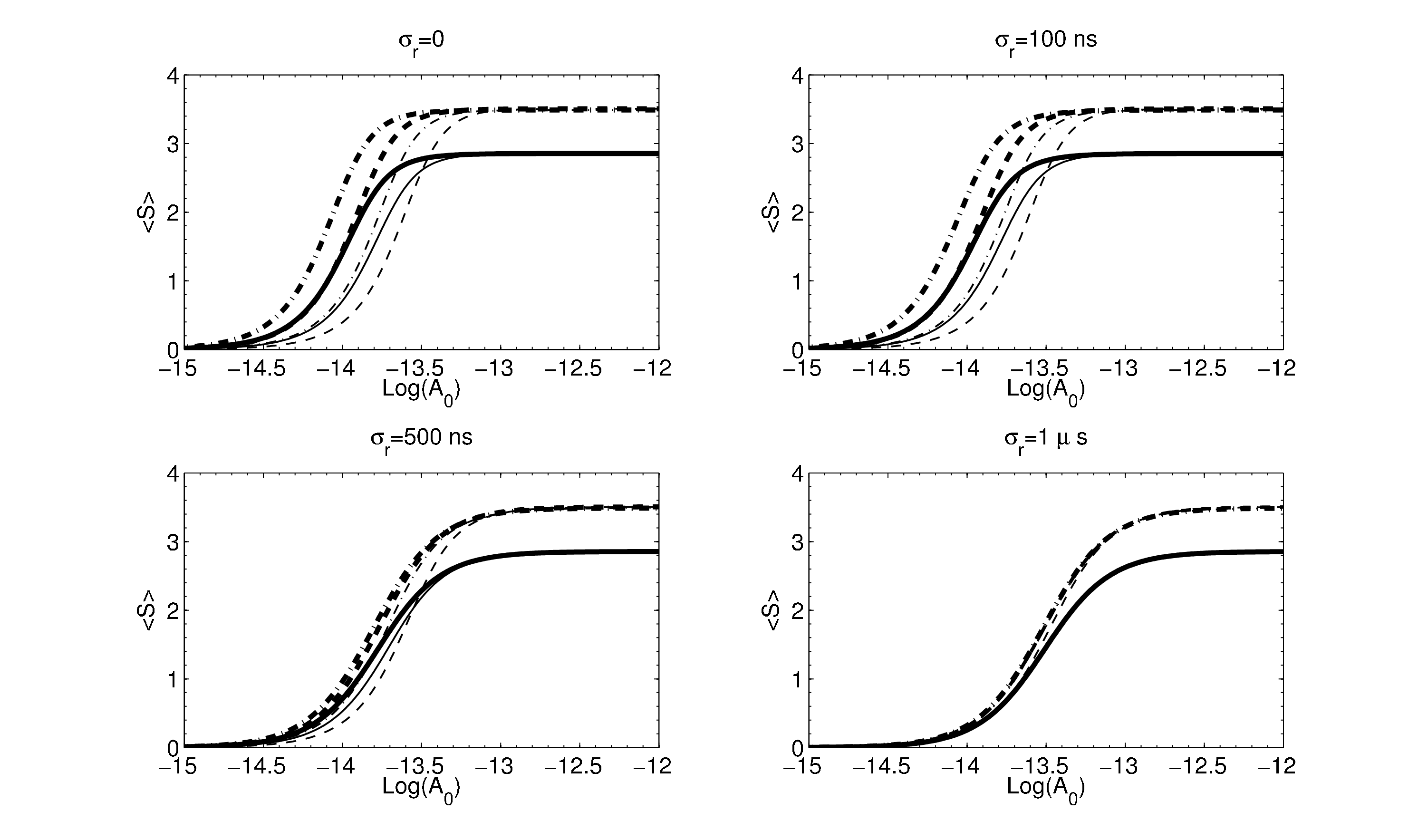}
\caption{ The GW background detection significance for PPTA, EPTA and
 NANOGrav with 5 year bi-weekly (one session every two weeks)
 observations. The white noise levels $\sigma_{0}$ are taken from
 Table~\ref{tab:psrp} and the red noise levels $\A \sigma_{\rm r}$
 are given above each panel. The x-axis shows the characteristic
 strain $A_0$ of the GW background with spectral index of
 $\alpha=-2/3$. The y-axis shows the expected detection significance
 $\langle S \rangle$. The solid lines, dashed lines, and
 dashed-dotted lines are for PPTA, EPTA and NANOGrav
 respectively. The thick lines are the optimized cases, while the
 thin lines are the unoptimized versions. Here, the constraint is the
 observation time, i.e.\,for each project, the total amount of
 observing time of each session is fixed to be 20 hours. If the red
 noise level is zero, the optimized array is able to detect 2-3 times
 weaker GW signals compared to its unoptimized version depending on
 the pulsar population.
 \label{fig:stw}
 }
 \end{figure*}

Figure~\ref{fig:stw} shows the comparison between the GW detection
significance $\langle S \rangle$ for an unoptimized and an optimized
PTA with the same parameters. We can see that the optimal observing
strategy increases the GW detection significance. Evaluating from the
rising edge of the curve, the optimized arrays are able to detect a GW
background 2-3 times weaker than unoptimized arrays depending on the
pulsar population. Because the radiometer noise $\sigma_{0}\gain^{-1}$
dominates over the pulse jitter noise for most pulsars using current
timing techniques, we ignore the pulse jitter noise in these examples.

We also note that the larger the red noise level is, the less we gain
from optimizing the observing schedule. When the amplitude of
intrinsic red noises is large, the $\A\B A$ dominates the detection
significance as shown in \EQ{eq:decse}, and thus the detection is no
longer sensitive to the schedule optimization, which only affects the
$\A \B B$ terms. In fact, only if the pulsar noise level is sensitive
to the observing schedule (i.e. when red noise does not dominate), the
optimization will be effective. This conclusion applies to any type of
GW detector.

In the optimization process, one always uses a numerical technique to
determine the optimal observation plan. It is, nevertheless, worth
finding a rule of thumb to determine an `approximate' optimal
strategy. We prove in \APP{sec:appopt} that, for strong GW cases, the
optimal observing schedule weakly depends on the amplitude of the GW
background and a good approximation for the optimal schedule is
\begin{equation}
 \A \tau=\tau\left(\frac{\sqrt{ \A Q} }{\sum_{j=1,\neq i}^{N} { \sqrt{\A Q} }
 }\right)\,,
 \label{eq:optapp}
\end{equation}
where the $\A Q$ is defined as
\begin{equation}
\A Q=\A \sigma_{0}^2 \gain^{-2} +\sigma_{\rm J}^2\,.
\end{equation}
Here the $\A Q$, noise parameter, defines the noisiness of the $i$-th
pulsar observed using a telescope of gain $\gain$.

\begin{figure}
 \centering \includegraphics[totalheight=2.5in]{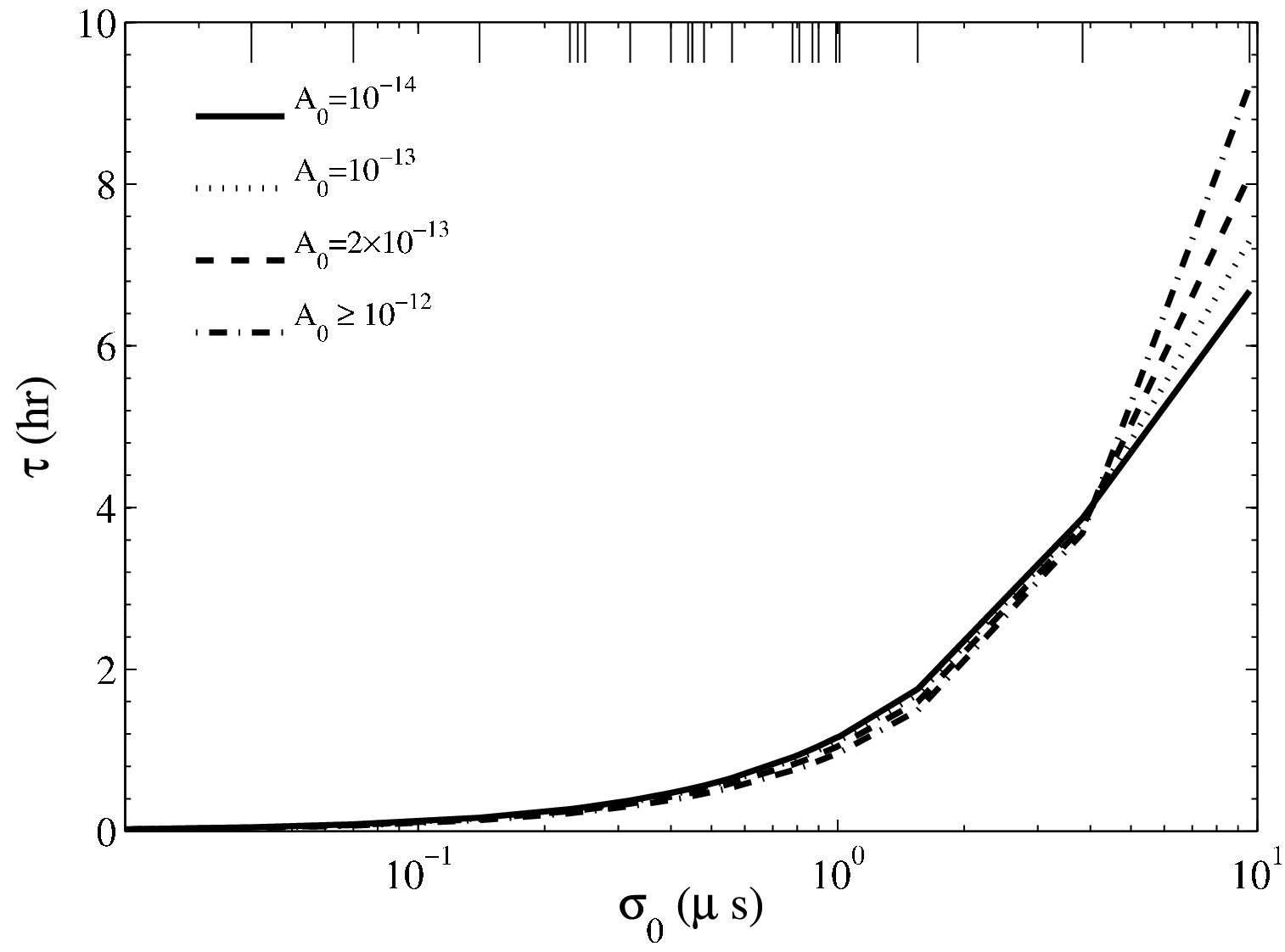}
 \caption{The optimal telescope time for each pulsar as a
  function of its white noise level. The x-axis shows the
  one-hour pulsar noise level $\A \sigma_{0}$. The y-axis shows
  the optimal telescope time for the pulsar per observing
  session. The EPTA pulsars are used in this demonstration,
  where the total telescope time is 24 hours, i.e. the average
	telescope time for each pulsar is 1 hour. We indicate $\A \sigma_0$ for each
  pulsar using a vertical line on the top of the figure. The
  dotted-dashed line, dashed line, dotted line and solid line
  are for optimization results with GW amplitude of
  $A_{0}=10^{-12}, 10^{-13}, 2\times 10^{-13}$, and $10^{-14}$
  respectively. The results of \EQ{eq:optapp} overlaps with the
  dotted-dashed line. For GWs with amplitude between $10^{-12}$
	and $10^{-13}$, the optimal schedules are very close to each
  other. For the weaker GW cases, e.g.~$A_{0}=10^{-14}$, the
  optimal schedule starts to deviate from the approximation
	\EQ{eq:optapp}. Such a deviation is mainly due to the pulsars
  with a high noise level. Since these noisy pulsars will not
  contribute significantly to the GW signal detection, the
  optimal algorithm starts to reduce its observing time. For
  most of the pulsars, the optimal schedule is still close to
  the result from \EQ{eq:optapp}. \label{fig:tauvssig}
} \end{figure}

By comparing the numerical optimization and the result from
\EQ{eq:optapp} in \FIG{fig:tauvssig}, we show that the optimal
schedule is insensitive to the GW amplitude and can be well
approximated by \EQ{eq:optapp}, when the amplitude of the GW is
large. For the case of a weak GW background, the optimal schedule for
most of the pulsars is still close to \EQ{eq:optapp}, but the optimal
schedule for noisy pulsars (pulsars with larger $\A Q$) starts to
deviate from the analytic approximation.

\begin{figure} \centering \includegraphics[totalheight=2.6in]{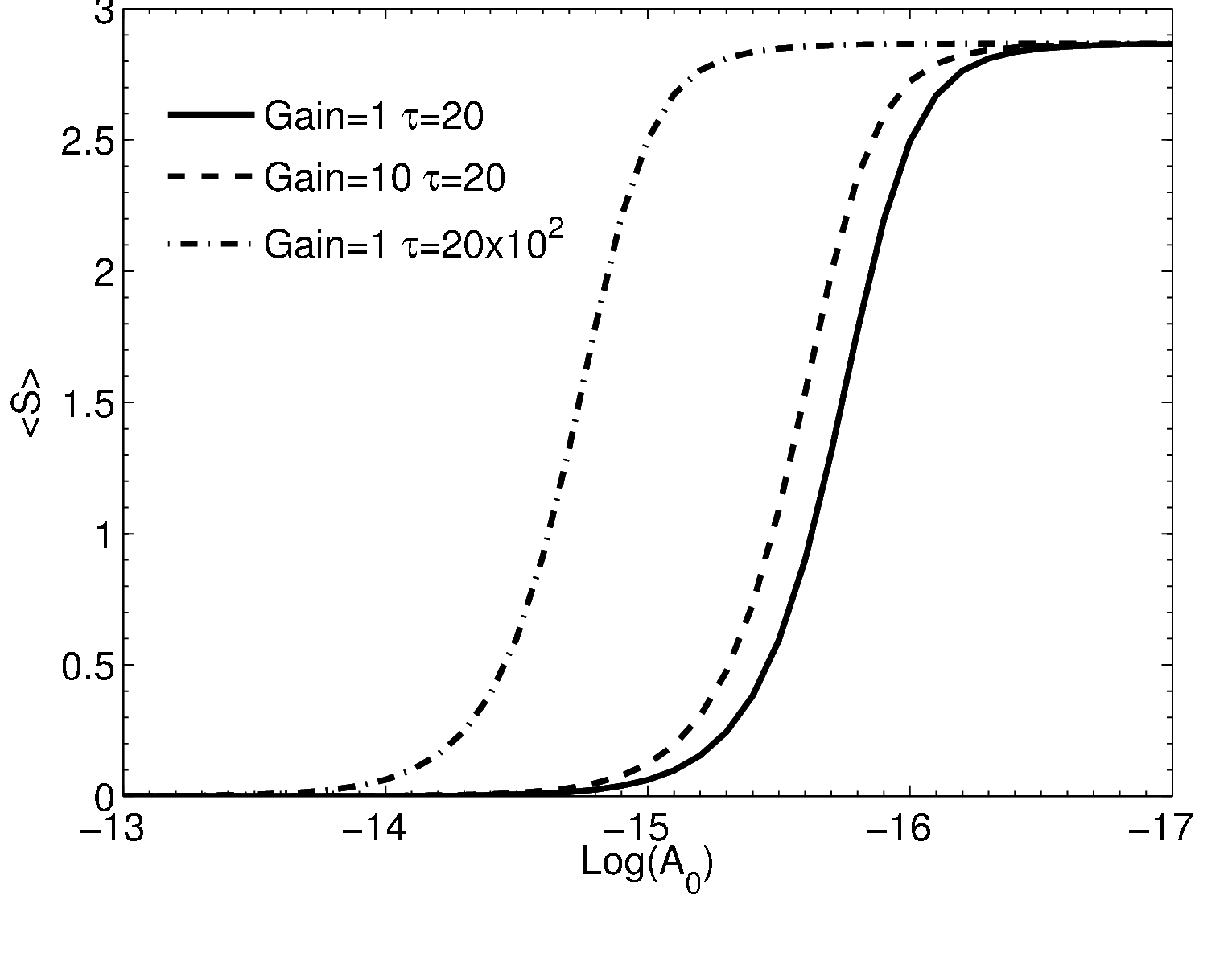} \caption{The
 GW detection significance $\langle S \rangle$ (y-axis) as a
 function of GW amplitude $\log(A_0)$ (x-axis), given different
 telescope gain and amount of telescope time. In the calculation, we
 assume that the PTA is composed of 20 pulsars, all of which have
 $\sigma_{0}=100$\,ns and $\sigma_{\rm J}=100$\,ns. We plot three
 telescope configurations here, each of which is labeled with the
 numerical value of total telescope time ${\tau}$ and telescope gain
 $\gain$. Here the telescope time is in units of hours, the gain
 takes an arbitrary fiducial unit as in \EQ{eq:scale}. For the
 following two PTA configurations, the $\{\gain=10, \tau=20\}$ and
 the $\{\gain=1, \tau=20\times10^2\}$, all pulsars have the same
 level of radiometer noise, which is 10 times smaller than the case
 where $\{\gain=1, \tau=1\}$. However, because of the pulse jitter
 noise effect, the configurations with more telescope time performs
 better than the cases with higher system gain. Furthermore, by
 comparing $\{\gain=10, \tau=20\}$ case with the case of $\{\gain=1,
 \tau=20\times 10^2\}$, we conclude that it is important, even for a
 high gain telescope, to acquire sufficient telescope time in order
 to improve the GW detection significance, when the pulsar becomes
 pulse jitter noise dominant.
 \label{fig:svsn} }
\end{figure}

There are a few more general conclusions on improving the timing accuracy
independent of GW detection algorithms. In order to improve timing
accuracy, one can use telescopes with higher effective
gain\footnote{Here, telescopes with higher effective gain $\gain$ can
 be telescopes with larger collection area, lower system temperature,
	wider bandwidth and so on. } or increase telescope time.
 Increasing the telescope gain reduces $\sigma_{\rm w}$ and
 increasing telescope time reduces both $\sigma_{\rm w}$ and $\sigma_{\rm J}$.  
 The example of case `D' in \FIG{fig:mulexam} shows
that one should use a low gain telescope on those pulsars with larger
jitter noise level, while using a high gain telescope on pulsars with
larger radiometer noise. This is further supported by the results in
\FIG{fig:svsn}, which shows that increasing the telescope time is more
effective than using a high gain telescope for pulse jitter noise
dominant pulsars.

Besides providing the optimal schedule to detect a GW background, our
technique answers the question about the optimal \emph{number} of
pulsars one should observe in a PTA with a given amount of telescope
time. The number of pulsars in a PTA has two effects on the GW
detection significance. Firstly, from \EQ{eq:decse}, one can see that
the significance increases as $\langle S\rangle \propto \sqrt{M}\sim
N$. Secondly, since the telescope time is limited, observing more
pulsars increases the TOA noise level, i.e. $\sigma_{\rm w}^2\propto
N^{-1}$ given a fixed amount of telescope time. When $N$ becomes
large, the two effects mentioned above cancel each other out, and the
detection significance becomes insensitive to $N$. In general, when
the number of pulsars ($N$) is small, the detection significance
increases with $N$. If all pulsars have the same noise level, the
detection significance saturates for large $N$, where the saturation
level is mainly determined by the available telescope time. In
practice, when trying to include more pulsars in a timing array, pulsars with a 
higher noise level will be inevitably included, such that the
detection significance will decrease for any GW detection
algorithm. In this way, observing more pulsars does not necessarily
help detecting the GW background, unless one gets more telescope
time. Given the telescope time, the number of pulsars, at which the GW
detection significance achieves its maximum, is the optimal number of
pulsars one should use in the PTA. We propose the following algorithm
to determine the best sample of pulsars to observe. 
\begin{enumerate} \item From a group of
 to-be-observed pulsars, choose the two pulsars with smallest noise
 levels, then optimize the schedule, and compute the GW detection
 significance. 
 \item Include one extra pulsar, optimize the
 schedule, compute the GW detection significance, loop over the rest
 of the pulsars and find the new pulsar leading to the largest
 detection significance. 
 \item If the GW detection significance
 increases by including the new pulsar, add this pulsar to the list,
 and repeat the second step for the rest of the pulsars, otherwise
 the optimal set of pulsars is already achieved.
\end{enumerate}

When pulsar surveys discover a new pulsar,
we can check whether it is worth including it in the PTA by
using the above algorithm, although, to start with, sufficient observations are 
still
necessary to measure the noise properties of this pulsar.

\section{Discussions and Conclusions}
\label{sec:con}

In this paper, we have investigated a technique to optimize the
allocation of telescope time in a pulsar timing array to maximize the
GW detection significance given a fixed amount of telescope time. This
is done in two steps. First, the GW detection significance using
a correlation detector is calculated analytically as a function
of the white noise and red noise level of each pulsar, i.e.\,
\EQ{eq:decse}. Secondly, the GW detection significance is optimized
under the constraints that the allocated telescope time is fixed. The
constrained optimization is converted to a corresponding constraint-free
optimization using the coordinate transformations of \EQ{eq:transforma} and
(\ref{eq:transform}). Finally, the optimization is solved numerically
using the downhill simplex algorithm. For characteristic PTAs, the
optimized arrays are able to detect a GW background 2-3 times weaker than
unoptimized arrays depending on the pulsar population.  Besides the single
telescope case, we also derive the optimization algorithm for multiple
telescopes. We also examine the links between the multiple and single
telescope optimization. We investigate the optimal number of pulsars to
observe for a given PTA, where the algorithm to construct the optimal
group of pulsars from candidates is also included.

In our optimization, the total telescope time $\tau$ and $\tau_{\nu}$
are input parameters, that need to be determined before optimizing the
schedule. When defining a PTA project, one can start with a reasonable
amount of telescope time, say 20 hours per each session/telescope,
optimize the schedule, calculate the detection significance $\langle
S\rangle$, check if the detection is sensitive enough to the predicted
GW background, and then adjust the input total telescope time
accordingly, i.e.\,increase total telescope time, if the PTA is not
sensitive enough.

In practice, the detailed numerical values for the optimal observing
strategy are still to-be-determined, due to the lack of measurements
for the necessary pulsar noise parameters. To determine the realistic
optimization strategy, these parameters are critical. A detailed
investigation on the individual timing properties of potential PTA
pulsars is highly necessary, from which further observations will
benefit.

One may encounter the situation, in which the optimal strategy requires
longer observing time for certain pulsars per session than their
transit time. If this happens, one has to split one observing session
into several successive sessions. The GW detection significance is not
significantly impaired by session splitting, since the GW induced
timing signal has a very red spectrum and we are detecting its low
frequency component. However the session splitting can lead to
inconveniences in practical arrangements. A better way to avoid such a
situation is to construct the PTA using telescopes with enough
geographical coverage, which can be one of the driving reasons for the
IPTA project.

The optimization in this paper is designed to maximize the GW
detection significance. Our optimization is built for the correlation
detector proposed by \cite{JHLM05}. As shown in \APP{sec:appopt}, our
optimization, which is very close to minimizing the total PTA noise
power, can be different from optimization for other types of GW
detectors e.g.~frequentist detectors \citep{VBC09, YCH11}
or the Bayesian detector \citep{VLM09, VLJ11}. In fact, as already
shown by \cite{BLF10}, one can get a different optimal observation
schedule, when focusing on single source detection. We have ignored
the information of historical non-overlapping data in our algorithms,
because the non-overlapping data has very limited contributions to the
cross-power of GW signals, although these data can be important to
constrain the upper limit of GW amplitude \citep{VLJ11}. Furthermore,
the PTA offers an opportunity to investigate much broader topics,
e.g.~interstellar medium effects, time metrology, and so on. This
paper, thus, by no means claims that our objective function is the
only one we should pursue. However our basic framework of
optimization, constraints and related numerical techniques will be the
same for other detectors. It is straightforward to generalize our
method to these detectors. For Bayesian detectors, there is no
analytical expression for the detection significance at present. The
Baysian detectors are usually computationally expensive, which makes
the optimization difficult at this stage.

Figure~\ref{fig:stw} shows that increasing the levels of red noise
does not decrease the saturation level of the detection significance,
i.e.\,the detection significance at large GW amplitude. This seems to
contradict the conclusion of \cite{JHLM05} and \cite{VBC09}, where the
red spectrum of the GW limits the saturation level. As shown in \EQ{eq:decse}, 
the term limiting the
saturation level is due to the term $(1+H(\A \B
\theta))\gamma_{kk'}^2$ in $\A\B A$, which is the non-zero correlation
between GW signals of two different pulsars. Since the red intrinsic
noise of pulsars, unlike the GW induced signal, are uncorrelated, it
is natural that they do not limit the saturation level.

In this paper, we use a phenomenological model to describe the noise
component, which is a superposition of a telescope-time dependent
white noise component and another red noise component independent of
telescope time. The reason for using such a phenomenological model is
to use the observational information and to introduce minimal
theoretical assumptions. Our white noise term contains both the
radiometer noise and pulse jitter noise. Although the radiometer noise
is the current bottleneck in timing accuracy of most MSPs (
$\sigma_{0} \gain^{-1}\gg \sigma_{\rm J}$), the pulse jitter noise
\citep{LK11}, can be a potential limitation for future single dish high gain
telescopes. Similarly, red noise can be another limitation for the
long term timing accuracy. Detailed studies on the pulse jitter noise
and red noise properties and related mitigation algorithms will be
useful for the future prospects of GW detection.

We assumed that the noise sources are not correlated among
pulsars. This may be valid for all the noise of astrophysical origin,
although the clock error can be an identical noise among all pulsars
\citep{HR84, FB90, Man94, Tinto11}. The clock error may also introduces 
correlations in TOAs from different telescopes, since observatory clocks are 
usually synchronized. However, thanks to the red
spectrum of most clock errors \citep{Riehle04}, one can completely
remove the noise using simultaneous differential measurements
\citep{Tinto11}. Furthermore, as we argue in \APP{sec:cnm}, such a
common noise source can be significantly suppressed by post
processing, if each session is compact within the time scale of days.
In this way, our assumption is justified that different noise sources
are not correlated among pulsars.

Our optimal observation strategy is for allocating telescope time among
pulsars. This only affects the white noise related part ($\A \B B$
terms) in GW detection significance (\EQ{eq:decse}). In fact, one can
also specify the epoch of each session to minimize the effect of the red
noise related part ($\A \B A$ terms). Since these $\A\B A$ terms can
be more effectively reduced using whitening techniques \citep{JHLM05,
JH06}, we do not optimize the epoch of each session to avoid sampling
artifacts and to reduce the complexities in observation management. The
discussions for the frequency of observation sessions are omitted,
since it is not bounded in terms of optimization, i.e. observing more
frequently sessions simply increase the sensitivity.

\section*{Acknowledgments}

We gratefully acknowledge support from the ERC Advanced Grant
`LEAP', Grant Agreement Number 227947 (PI Michael Kramer). We thank
Norbert Wex for reading through the manuscript and his useful
discussions. We also thank the anonymous referee for the important and helpful 
comments.

\bibliographystyle{mn2e}

\clearpage

\appendix

\section{Pulsar timing correlation}
\label{sec:appcor}
In this section, we calculate the cross correlation for a Gaussian
random signal with a power-law spectrum. We discuss two main issues
arising in the calculation, the effect of polynomial fitting on the
cross correlation and the effects of spectral leakage. For a stationary
continuous-in-time random signal $s(t)$, which has a prescribed power
spectrum of $S_{\rm s}(f)$, the well-known Wiener-Khinchin theorem
states that the autocorrelation $\langle s(t_{i}) s(t_{j}) \rangle$ is
the Fourier transform of the power spectral density,
i.e. \begin{equation} \langle s(t_{i}) s(t_{j})
 \rangle=\int_{0}^{\infty}S_{\rm s}(f) e^{2\pi f (t_{i}-t_{j})}\,
 df\,.
 \label{eq:wctheor} \end{equation} However due to the fitting of polynomials,
 the direct application of the above to the pulsar timing problem needs 
 revision.

\begin{figure} \centering
 \includegraphics[totalheight=2.5in]{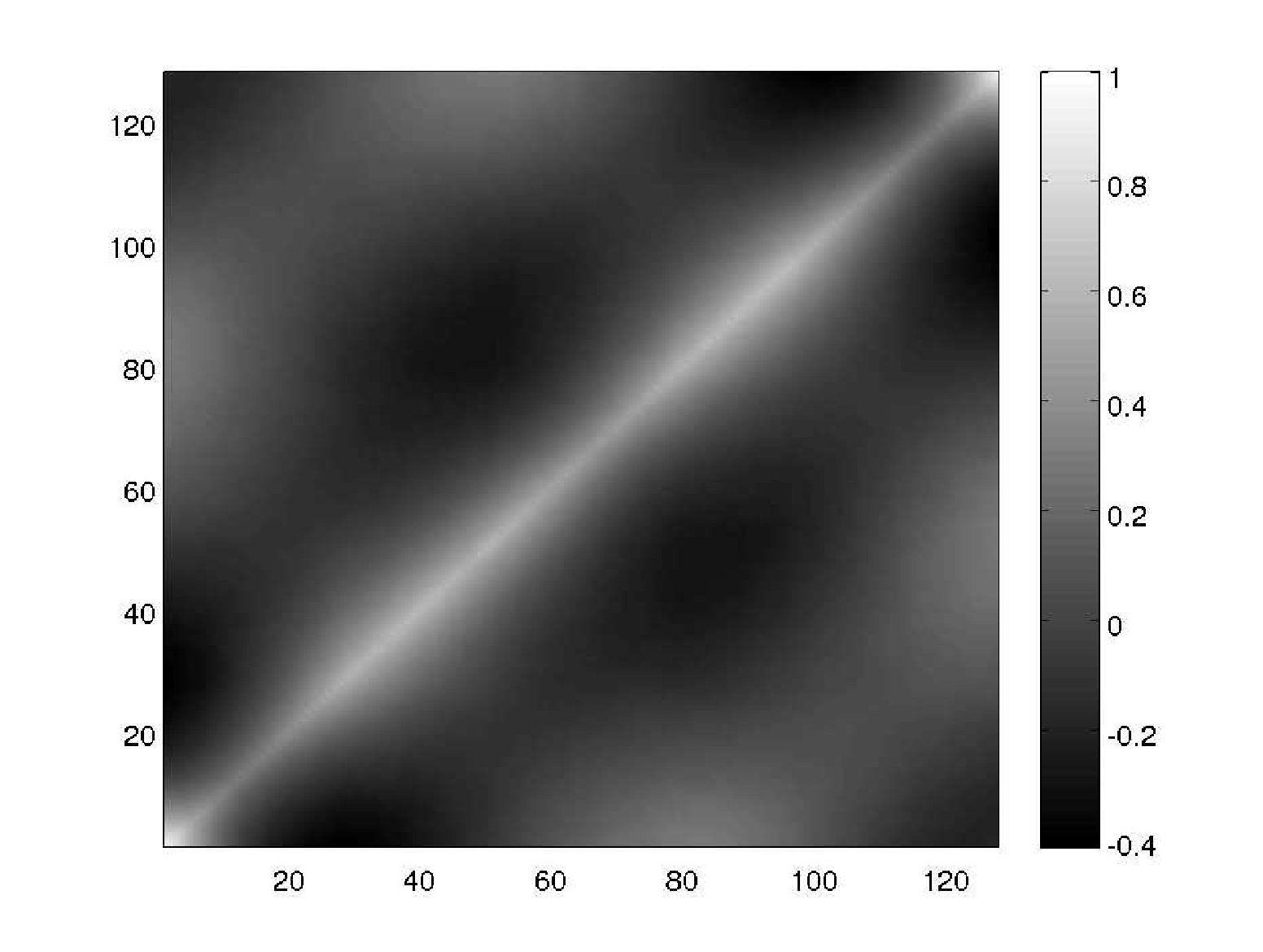}
 \caption{ The correlation coefficient for 128 regular samples
  of a red noise power-law. For illustration purposes the
  spectral index is -2 and we removed the fitted parabolic term
  in the signal $s(t_{i})$. The x-axis and y-axis show the
  index $i$ and $j$, respectively. The colors indicate values
  of the correlation coefficients. The diagonal elements of the
  plot are not equal to each other. This clearly show that the
  polynomial fitting breaks the stationarity of the red
  noise. \label{fig:coreff}} \end{figure}

In a practical pulsar timing data reduction pipeline, one usually uses
a least-square polynomial fitting to extract the pulsar
parameters. Such a fit is \emph{not} a stationary process, which
prevents us from calculating the correlation directly using
\EQ{eq:wctheor}. In this paper, correlations are calculated via
numerical simulations, that take the following steps: i) Generate a
series of the sampled signal using individual frequency components as
described in \cite{LJR08} ii) Fit the signal with a polynomial to
simulate the effects of fitting the pulsar rotation frequency and its
derivative. iii) Calculate the required cross correlation. iv) Repeat
steps (i), (ii), (iii) and average the cross correlation, until the
required precision is attained. In this paper, the correlations are
calculated to a relative error of 1\%. From the numerical results in
\FIG{fig:coreff}, the polynomial fitting clearly breaks the
stationarity of signals. 

The other issue is the spectral leakage. A red noise signal with a
steep spectrum introduces leakage from low frequency components into
high frequency components. For a red noise signal with a spectral
density of $S_{\rm s}(f)\sim f^{-\beta}$, the amplitude for those
signal components within the frequency range $[f, f+\delta f]$ is
$f^{-\beta/2} \delta f^{1/2}$. The waveform of this component will be
sinusoidal, i.e.\,$s(t)\simeq f^{-\beta/2}\delta f^{1/2} \exp(2\pi f i
t)$. At the low frequency limit, where $f t \ll 1$, $s(t)$ can be
approximated using Taylor series
\begin{equation}
 s(t)\sim f^{-\beta/2} \delta f^{1/2} \exp(2\pi i f t)\simeq \delta f^{1/2}
 \sum_{l=0}^{l=\infty} \frac{(2\pi i t)^{l} f^{l-\beta/2}}{l!}.
 \end{equation}
If we fit the waveform with a $n$ degree polynomial, we effectively
remove the leading terms up to $f^{n+3/2-\beta/2}$. Clearly, if
$n+3/2-\beta/2\le0$, the signal amplitude goes to infinity, when
$f\rightarrow 0$. Thus there will be no low frequency cut-off in the
spectrum, even for data with a finite length. In this case, the low
frequency components are always dominant. To guarantee that the low
frequency cut-off rises naturally (regularize the signal), we need to
fit the time series to a polynomial with degree $n > \beta/2-3/2$.
For example, the SMBH GW background introduces a pulsar timing signal
with a spectral index of $\beta=13/3$, hence we need at least $n=1$,
i.e.\,fitting with a first order polynomial (a straight line),
although the spectral leakage from low frequency components is
dominant.

We now take a closer look at the effects of the fitting on the RMS
level of a signal with a power-law spectrum given by
\begin{equation}
 S_{\rm s}(f)=S_0 f^{-\beta}, \textrm{with }f\in[f_{\rm L}, \infty)\,,
\end{equation}
where the $f_{\rm L}$ is the lowest frequency cut-off, $S_0$ is the power 
spectrum value at the unit frequency.  Using the data
length ($T$) as the temporal unit, the RMS level of such signal is
\begin{equation}
 \sigma^2= S_0 T^{\beta-1}\frac{f_{\rm 0,L}^{1-\beta}}{ (\beta-1)}\,,
 \label{eq:refz}
\end{equation}
where $f_{\rm 0, L}=T f_{\rm L}$ is the dimensionless lowest cut-off
frequency.

The polynomial fitting to regular sampled data using a $\ell^2$-norm is 
equivalent to the
following \emph{linear time-variant} filter
\begin{equation}
 h(t_1,t_2)=\sum_{l=0}^{n}(2l+1)P_l(2t_1-1)P_l(2t_2-1)\,,
\label{eq:polfit}
\end{equation}
where $t_1,t_2 \in [0,1]$ are dimensionless time, $n$ is the order of
the fitting polynomial, and $P_{l}(\cdot)$ is the $l$-th order
Legendre polynomial. Denoting the signal as $s(t)$ and the polynomial
fitting of such a signal as $s_0(t)$, we have \begin{equation}
 s_0(t)=\int_0^1 h(t,t_2) s(t_2)\,\D t_2\,,
\end{equation}
and the residual is $s'(t)=s(t)-s_0(t)$.

The average RMS level of post-fitting residual becomes
\begin{equation}
 \sigma'^2=\int_{0}^{1} \D t s'(t)^2 =\int_{0}^1\int_0^1\int_0^1
 h(t,t_1)h(t,t_2) s(t_1)s(t_2)\, \D t \D t_1 \D t_2\,.\end{equation}
One can show that
\begin{equation}
 \sigma'^2=\sigma^2- \int_0^1\int_0^1 \sum_{l=0}^{n} (2l+1)
 P_l(2t_1-1)P_l(2t_2-1) C(t_1-t_2)\, \D t_1 \D t_2\,,
 \label{eq:int}
\end{equation}
where $C(t_1-t_2)=\langle s(t_1)s(t_2)\rangle$. For the case of a power-law
spectrum, we have
\begin{eqnarray}
 C(t)&=&S_0 T^{\beta-1}\frac{f_{\rm 0, L}^{1-\beta}}{\beta-1}
 \,_1F_2\left(\frac{1-\beta}{2}; \frac{1}{2}, \frac{3-\beta}{2};
 -\pi^2 f _{\rm 0, L}^2 t^2\right)\nonumber \\ &&+
 T^{\beta-1}(2\pi)^{\beta-1} \Gamma(1-\beta)
 \sin\left(\frac{\pi\beta}{2}\right)|t|^{\beta-1}\,,
\end{eqnarray}
where $\tau$ is dimensionless time in units of $T$, $\,_1F_2(\cdot)$
is the generalized hypergeometric function (see also \cite{VLJ11} for
the series presentation), and $\Gamma(\cdot)$ is the gamma function.

Integrating \EQ{eq:int}, one reads \begin{eqnarray}
 \sigma'^2&=&S_0 T^{\beta-1}\left[\frac{}{} \right.\nonumber
 \\ &-&\sum_{k=n+1}^{\infty} \frac{(2\pi)^{2k} (-1)^{k+n} f_{\rm
 0,L}^{2k+1-\beta}(1+n) \Gamma(k) \sin(\pi \beta)
 }{(1+2k-\beta)\Gamma(2+2k)\Gamma(k-n) \Gamma(2+k+n)} \nonumber
 \\ &+&\left.2^{\beta-2}(1+n)\pi^{\beta-1}
 \frac{\Gamma(\frac{3+2m-\beta}{2})
 \Gamma(\frac{\beta-1}{2})\Gamma(\frac{1+\beta}{2})}{
 \Gamma(\frac{3+2n+\beta}{2}) \Gamma(1+\beta)}\right]\,.
\end{eqnarray}
Clearly, if $2n+3\ge\beta$, we have
\begin{equation}
 \lim_{f_{\rm 0, L}\rightarrow0}
 \sigma'^2=S_0 T^{\beta-1}2^{\beta-2}(1+n)\pi^{\beta-1}
 \frac{\Gamma(\frac{3+2n-\beta}{2})
 \Gamma(\frac{\beta-1}{2})\Gamma(\frac{1+\beta}{2})}{
 \Gamma(\frac{3+2n+\beta}{2}) \Gamma(1+\beta)}\,,
\end{equation}
i.e. the RMS $\sigma'^2$ is regularized to be a finite value for
$f_{\rm 0,L} \rightarrow 0$, which confirms our previous
estimation. Usually pulsar spin and spin-down are subtracted from the data. This 
corresponds to the case of $n=2$.  For GW induced signal, we have $\beta=13/3$ 
and $S_0=h_{\rm c}^2(\textrm{yr}^{-1})(12\pi^2)^{-1} \textrm{yr}^{-4/3}$, which 
gives
\begin{eqnarray}
	\sigma'^{2}&=&h_{\rm c}^2(\textrm{yr}^{-1})\frac{3^6 (2\pi)^{4/3}T^{10/3} 
	\Gamma\left(\frac{2}{3}\right)}{2^7 \cdot 5\cdot 7^2\cdot 11\cdot 13} \,
	\textrm{yr}^{-4/3}\\
	&\simeq&2.55\times 10^{-3} h_{\rm c}^2T^{10/3}\, \textrm{yr}^{-4/3}\,.
\end{eqnarray}
This is identical to the results of \cite{VL12}. 

It should be also noted that for such case ($\beta=13/3$ and $S_0=h_{\rm 
c}^2(\textrm{yr}^{-1})(12\pi^2)^{-1} \,\textrm{yr}^{-4/3}$) \EQ{eq:refz} 
becomes \begin{equation}
	\sigma^2\simeq2.53\times 10^{-3} h_{\rm c}^2T^{10/3} f_{\rm 0, L}^{-10/3}\, 
	\textrm{yr}^{-4/3}\,.
\end{equation}
In this way, estimating RMS value of fitted signal using \EQ{eq:refz} is 
accurate enough for practical purposes, given data length is adopted as the 
\emph{effective cut-off frequency}, i.e. to use $f_{\rm 0,L}=1$. 

\section{GW detection significance}
\label{sec:appS}

In this section, we calculate the GW detection significance. The
expected value for the detection significance $\langle S \rangle$
depends on $\Sigma_{\rm c}$ and $\langle \A \B c \rangle$ as shown in
\EQ{eq:exps}. We determine the $\langle \A \B c \rangle$ first. From
\EQ{eq:defcorr}, we have \begin{equation} \langle \A \B c \rangle=
 {\frac{1}{m}\sumtim \langle \Ra_{k} \Rb_k \rangle }= {\sigma_{\rm
 g}^2 H(\A \B \theta) }\,. \label{eq:expc}
\end{equation}

To determine $\Sigma_{\rm c}$, we need $\langle \A \B c^2\rangle$,
which is calculated in a similar fashion such that \begin{equation} \langle
 \A \B c^2\rangle={\frac{1}{m^2}\left\langle \sum_{k=1}^{m}
 \sum_{k'=1}^{m} \A R_{k} \B R_{k} \A R_{k'} \B R_{k'}\right\rangle
 }\,, \label{eq:avrc2}
\end{equation} After using the correlation relation \EQ{eq:corgws},
\EQ{eq:cornoi}, and performing the Wick expansion \citep{Zee10} to
calculate higher momentum, we have \begin{equation} \left \langle
 \frac{1}{m^2}\sum_{k=1}^{m} \sum_{k'=1}^{m} \A R_{k} \B R_{k} \A
 R_{k'} \B R_{k'}\right \rangle = \sigma_{\rm g}^4 \left( \A\B A+\A\B
 B+H(\A \B \theta)^2 \right ) \,. \label{eq:rrrr}
\end{equation} The $\Sigma_{\rm c}$ is then \begin{eqnarray}
 \Sigma_{\rm c}&=&\sqrt{\left \langle {\frac{1}{M}\sum_{\ijp} \left(\A\B
 c-\overline{\langle c\rangle}\right)^2} \right
 \rangle}\nonumber \\
&=& \sqrt{\frac{1}{M}\sum_{\ijp} \left(\langle \A \B c ^2 \rangle
-\overline{\langle c\rangle}^2\right)} \nonumber \\
&=&{\sigma_{\rm
g}^2}\sqrt{\Sigma_{H}^2+\frac{1}{M}\sum_{\ijp} \left(
\A\B A+\A\B B\right)}\,, \end{eqnarray} with which one can derive
\EQ{eq:decse}.

\section{Optimization using Lagrangian multiplier}
\label{sec:appopt}
In section~\ref{sec:optintro}, we describe the optimization technique
using variable transformation and numerical optimization. In this
section, we introduce another method solving the constrained optimization
problem directly. As an example, we present the technique for the single
telescope situation, which is also readily generalized.

The optimization problem for the single telescope case is to search
the minimal value of $\Lag=\sum_{\ijp} \AB B$ under the constraint
that $\tau=\sum_{i=1}^{n} \A \tau$. Using a Lagrangian multiplier
$\lambda$, we can re-cast the optimization problem to optimize the
$\Lag'$, where
\begin{eqnarray}
 \Lag'&=&\Lag+\lambda \left (\tau-\sum_{i=1}^{n} \A \tau\right) \\
 &=&\sum_{\ijp} \frac{\A \kappa \B q}{\B \tau} + \frac{\B \kappa \A q}{\A
 \tau}+ \frac{\A q \B q}{\A \tau \B \tau} +\lambda \left (\tau-\sum_{i=1}^{n}
 \A \tau\right)\,, \label{eq:lagequiv}
\end{eqnarray}
where $\A q=(\A \sigma_{0}^2\gain^{-2} + \A \sigma_{\rm
 J}^2)\sigma_{\rm g}^{-2}$ and $\A \kappa =1+\A \sigma_{\rm r}^2
\sigma_{\rm g}^{-2}$. The minimization of $\Lag'$ can be found by
\begin{eqnarray}
 \frac{\partial \Lag'}{\partial \A \tau}&=&0\,, \label{eq:diffe1}\\
 \frac{\partial \Lag'}{\partial \lambda}&=&0\,, \label{eq:diffe2}\end{eqnarray}
which give
\begin{equation}
 \frac{\A q}{\A \tau^2}\sum_{j=1, \neq i}^{n}\B \kappa+\frac{\A
 q}{\A \tau^2}\sum_{j=1, \neq i}^{n} \frac{\B q}{\B
 \tau}-\frac{\lambda}{2} =0 \,,\\
 \label{eq:difopt1}
\end{equation}
and \begin{equation}
 \sum_{i=1}^{n} \A \tau-\tau =0\,.
 \label{eq:difopt2}
\end{equation}
It is easy to check that \EQ{eq:difopt1} and ~(\ref{eq:difopt2}) can
be solved using the following recipe:

1. Guess an initial value for $\A \tau$.

2. Update $\A\tau$ with a newer value using

\begin{equation}
 \tau \frac{\sqrt{ \A q\sum_{j=1,\neq i}^{n}\left(\B \kappa+{\B
 q}/{\B\tau}\right)}} { \sum_{i=1}^{n} \sqrt{\A q\sum_{j=1,\neq i}^{n}\left(\B
 \kappa+{\B q}/{\B\tau}\right)}} \rightarrow \A\tau
 \label{eq:iteration}
\end{equation}

3. Repeat step 2, until the required precision is achieved.

One can monitor the change of $\A\tau$ for each iteration until it
converges to the necessary precision. The initial value for the
iteration is determined from the strong-signal limit, i.e.\,$\A
q\rightarrow 0$, where the iteration \EQ{eq:iteration} reduces to a
solution \begin{equation} \A \tau=\tau \frac{\sqrt{\A q \sum_{j=1,\neq
 i}^{n} \B \kappa}}{\sum_{i=1}^{n} \sqrt{ \A q \sum_{j=1,\neq
 i}^{n} \B \kappa}}+{\cal O}(q)\,.
 \label{eq:initialv}
\end{equation}

It is worthwile noting that, in fact, the iteration process
(\EQ{eq:iteration}) will not change the results very much from the
initial value \EQ{eq:initialv}. This is due to

\begin{eqnarray}
 &&\tau \frac{\sqrt{ \A q\sum_{j=1,\neq i}^{n}\left(\B \kappa+{\B
 q}/{\B\tau}\right)}} { \sum_{i=1}^{n} \sqrt{\A q\sum_{j=1,\neq i}^{n}\left(\B
 \kappa+{\B q}/{\B\tau}\right)}} \nonumber \\
 &&\simeq \tau \frac{\sqrt{ \A q\sum_{j=1}^{n}\left(\B \kappa+{\B
 q}/{\B\tau}\right)}} { \sum_{i=1}^{n} \sqrt{\A q\sum_{j=1}^{n}\left(\B
 \kappa+{\B q}/{\B\tau}\right)}}+{\cal O}\left(\frac{1}{n}\right) \nonumber \\
 &&=\tau \frac{\sqrt{\A q}}{\sum_{i=1}^{n} \sqrt{\A q}} \nonumber \\
 &&= \tau \frac{\sqrt{\A Q}}{\sum_{i=1}^{n} \sqrt{\A Q}} \,,\end{eqnarray}
 where $\A Q=\A \sigma_{0}^2 \gain^{-2} +\sigma_{\rm J}^2$. Clearly, the
 initial value \EQ{eq:initialv} we use is already a good approximation to the
 optimal observation strategy.

\section{The condition of existing solutions to $\A P_{\nu}$}
\label{app:exist}
In this section, we investigate the conditions to be satisfied such
that the following equations have solution $\A P_{\nu}$ for given $\A
\tau_{\rm e}, \gain_{\nu}, \A O_{\nu}$, and $\tau_{\nu}$,
\begin{empheq}[left=\empheqlbrace]{align}
 \A\tau_{\rm e}&=\sum_{\nu=1}^{N_{\rm tel}} \gain_{\nu}^2 \A P_{\nu}
 \A O_{\nu}\,, \label{eq:a1} \\ \tau_{\nu} &=\sum_{i=1}^{N}\A
 P_{\nu} \A O_{\nu}\,, \label{eq:a2}\\ \A
 P_{\nu}&\ge0\,, \label{eq:a3}\end{empheq} where $i$ is $1\ldots N$,
$\nu$ is $1\ldots N_{\rm tel}$, and $\A \tau_{\rm e}$ satisfies
\EQ{eq:taueff}.

We want to prove that \EQ{eq:a1}, (\ref{eq:a2}), and (\ref{eq:a3})
have solutions $\A P_{\nu}$, if and only if, for any $i$, the
following condition is satisfied, \begin{equation} \A \tau_{\rm e} \le
 \sum_{\nu=1}^{N_{\rm tel}} \gain_{\nu}^2 \tau_{\nu} \A O_{\nu}\,.
 \label{eq:cond}
\end{equation}

The proof contains two steps as follows: i) The `only if' part. If
$\A P_{\nu}$ is the solution, then
\begin{equation}
 \A \tau_{\rm e}=\sum_{\nu=1}^{N_{\rm tel}} \gain^2_{\nu} \A P_{\nu} \A
 O_{\nu}\le
\sum_{\nu=1}^{N_{\rm tel}} \gain^2_{\nu} \tau_{\nu} \A O_{\nu}\,,
 \label{eq:provof}
\end{equation}
where the second step is due to the constraint of \EQ{eq:a2} and
(\ref{eq:a3}). ii) The `if' part. It is easy to note that, under the
constraints of \EQ{eq:a2}, any linear combination of the column
vectors, e.g.~$\beta_i=\sum_{\nu} c_{\nu} \A P_{\nu} \A O_{\nu}$,
belongs to a hyperplane $\cal S$, because $\sum_{i=1}^{N}
\beta_i=\sum_{\nu} \tau_{\nu} c_{\nu}$. Due to \EQ{eq:a3}, only part
of the hyperplane $\cal S$ is accessible to the vector $\beta_{i}$,
i.e.\,$\beta_{i}$ is constrained by $0\le \beta_{i}\le \sum_{\nu}
c_{\nu} \tau_{\nu} \A O_{\nu}$ in the $\cal S$. Denoting such an
accessible region as $\cal A$, and let
$c_{\nu}=\gain^{2}_{\nu}$. Clearly, if the vector $\A \tau_{\rm e}$ is
in the accessible region $\cal A$, there exists a solution to
\EQ{eq:a1}, which satisfies both \EQ{eq:a2} and (\ref{eq:a3}). From
\EQ{eq:taueff}, we know $0\le\A \tau_{\rm e} \le \sum_{\nu=1}^{N_{\rm
 tel}} \gain_{\nu}^2 \tau_{\nu} \A O_{\nu}$, thus the vector $\A
\tau_{\rm e}$ belongs to the accessible region $\cal A$, and the
solution exists.

\begin{figure} \centering \includegraphics[totalheight=2.5in]{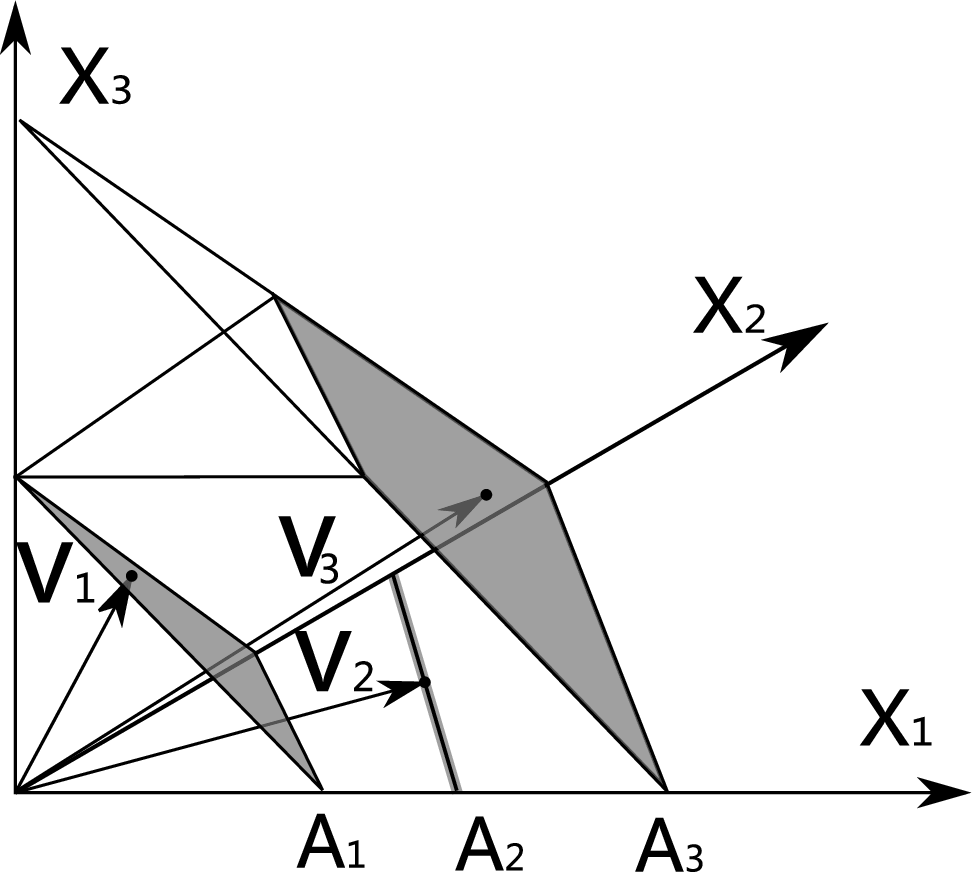}
\caption{The illustration for the accessible region of the summation of
vectors. Here $X_1, X_2, X_3$ are the coordinates. Because of the $\A
O_{\nu}$ we choose, the vector $v_1$ is constrained to the region $A_1$
(the plane constrained to the shaded triangle) and the vector $v_2$
is constrained to the region $A_2$ (the line segment). The accessible
region for the summation ($v_3=v_1+v_2$) of two vectors is clearly the
$A_3$, i.e.\,a plane with extra constraints.
 \label{fig:illu}}

\end{figure}

A graphical illustration for the condition is given in \FIG{fig:illu},
where we choose $\A O_{\nu}=\left[\begin{array}{c c}1 & 1\\1& 1 \\ 1&
 0\end{array}\right]$.

\section{Common noise mitigation}
\label{sec:cnm}

Clock errors and other common noise sources are harmful to PTA
observations. \cite{Tinto11} has shown that a simultaneous
differential measurement of a pulsar TOA completely removes the clock
error. Instead of the requirement for simultaneous observations of
multiple pulsars, we argue in this section that one can still remove
most of the clock errors in post processing without simultaneity, if
each observation session is compact enough.

Suppose the pulsar timing signal contains an identical (clock) noise
$n_{\rm c}(t)$ with a red spectrum. A subtraction between the timing
signals of two different pulsars at two close epochs will remove most
power of the identical noise component. One can check this by looking
at the residuals (i.e.\,$n_{\rm c}(t)-n_{\rm c}(t+\Delta)$) of the
identical noise after the subtraction. It is easy to show that the
power spectrum of the residual $S_{\rm n}(f)$ becomes
\begin{equation}
 S_{\rm n}(f)=S_{\rm c}(f) [1-\cos(2\pi f\Delta)]\,,
 \label{eq:sigres}
\end{equation}
where $S_{\rm c}(f)$ is the noise spectrum of $n_{\rm c}$, $\Delta$ is
the time difference between the two epochs and is thus roughly the
time span of one observing session. The clock noise $S_{\rm c}(f)$
dominates at low frequencies with time scales of ten years, thus we
can check the residual at such frequencies, where the residual power
spectrum becomes
\begin{equation}
S_{\rm n}(f)\simeq \left[ 1.5\times 10^{-6} \left(\frac{f}{\rm 10\,
yr^{-1}}\right)^2 \left(\frac{\Delta}{\rm 1 \, day}\right)^2
\right]S_{\rm c} (f)
\end{equation}
Thus, if we keep each observation session compact within a few days,
the clock error can still be significantly (factor of $\sim 10^{6}$)
suppressed by post processing, and we do not need to worry about such
common noise for planning an observing schedule at this stage.

\label{lastpage}

\end{document}